\documentclass[twocolumn]{aastex631} %,linenumbers

\usepackage{mathptmx}
\usepackage[T1]{fontenc}
\usepackage{ae,aecompl}
\usepackage{graphicx}	
\graphicspath{{./Figures/}} 
\usepackage{amsmath}	
\usepackage{amssymb}	
\usepackage{subfigmat}
\usepackage{enumerate}
\usepackage{multirow}
\usepackage{natbib}

\newcommand{\kms}{\hbox{km s$^{-1}$}}

\newcommand{\funit}{\mbox{mJy~beam$^{-1}$}}

\usepackage{tabularx, booktabs}
\usepackage{rotating}
\usepackage{longtable}
\usepackage{ltablex}
\usepackage{threeparttablex}

\def\gtrsim{\mathrel{\hbox{\rlap{\hbox{\lower5pt\hbox{$\sim$}}}\hbox{$>$}}}}

\newcommand{\msun}{\mbox{M$_\odot$}}

\newcommand{\degree}{\mbox{$^\circ$}}

\newcolumntype{Y}{>{\centering\arraybackslash}X}

\shorttitle{Emission Distribution in V883 Ori}
\shortauthors{Lee et al.}

\begin{document}

\title{ALMA Spectral Survey of An eruptive Young star, V883 Ori (ASSAY): I. What triggered the current episode of eruption?}

\correspondingauthor{Jeong-Eun Lee}
\email{lee.jeongeun@snu.ac.kr}

\author[0000-0003-3119-2087]{Jeong-Eun Lee}
\affil{Department of Physics and Astronomy, Seoul National University, 1 Gwanak-ro, Gwanak-gu, Seoul 08826, Korea}
\affil{SNU Astronomy Research Center, Seoul National University, 1 Gwanak-ro, Gwanak-gu, Seoul 08826, Republic of Korea}

\author{Chul-Hwan Kim}
\affil{Department of Physics and Astronomy, Seoul National University, 1 Gwanak-ro, Gwanak-gu, Seoul 08826, Korea}

\author{Seokho Lee}
\affiliation{Korea Astronomy and Space Science Institute, 776 Daedeok-daero, Yuseong, Daejeon 34055, Korea}

\author[0000-0001-6324-8482]{Seonjae Lee}
\affil{Department of Physics and Astronomy, Seoul National University, 1 Gwanak-ro, Gwanak-gu, Seoul 08826, Korea}

\author{Giseon Baek}
\affil{Department of Physics and Astronomy, Seoul National University, 1 Gwanak-ro, Gwanak-gu, Seoul 08826, Korea}

\author{Hyeong-Sik Yun}
\affiliation{Korea Astronomy and Space Science Institute, 776 Daedeok-daero, Yuseong, Daejeon 34055, Korea}

\author{Yuri Aikawa}
\affiliation{Department of Astronomy, University of Tokyo, 7-3-1 Hongo, Bunkyo-ku, Tokyo 113-0033, Japan}

\author[0000-0002-6773-459X]{Doug Johnstone}
\affiliation{NRC Herzberg Astronomy and Astrophysics, 5071 West Saanich Road, Victoria, BC, V9E 2E7, Canada}
\affiliation{Department of Physics and Astronomy, University of Victoria, 3800 Finnerty Road, Elliot Building, Victoria, BC, V8P 5C2, Canada}

\author{Gregory J. Herczeg}
\affiliation{Kavli Institute for Astronomy and Astrophysics, Peking University, Yiheyuan 5, Haidian Qu, 100871 Beijing, China}
\affiliation{Department of Astronomy, Peking University, Yiheyuan 5, Haidian Qu, 100871 Beijing, China}

\author{Lucas Cieza}
\affiliation{Facultad de Ingenier\'ia y Ciencias, N\'ucleo de Astronom\'ia, Universidad Diego Portales, Av. Ejercito 441. Santiago, Chile}

\begin{abstract}
An unbiased spectral survey of V883 Ori, an eruptive young star, was carried out with the Atacama Large Millimeter/submillimeter Array (ALMA) in Band 6.
The detected line emission from various molecules reveals morphological/kinematical features in both the Keplerian disk and the  infalling envelope. A direct infall signature, red-shifted absorption against continuum, has been detected in CO, HCO$^+$, HCN, HNC, and H$_2$CO. HCO$^+$ and SO show large arm-like structures that probably connect from an infalling envelope to the disk. HCN and H$_2$CO reveal a distinct boundary between the inner and outer disk and reveal tentative spiral structures connecting the outer disk to the inner disk. HNC shows a large central emission hole (r $\sim$~0.3\arcsec) due to its chemical conversion to HCN at high temperatures. The HDO emission, a direct tracer of the water sublimation region, has been detected in the disk. Molecular emission from complex organic molecules (COMs) is confined within the HDO emission boundary, and HCO$^+$ has an emission hole 
in its distribution due to its destruction by water. Together, these features suggest that the current episode of eruption in V883 Ori may be triggered by the infall from the envelope to the outer disk, generating a spiral wave, which propagates inward and greatly enhances the accretion onto the central star. 
\end{abstract}

\section{Introduction}
The episodic accretion process \citep{hartmann96} is now accepted as the standard accretion model in low-mass star formation. Although we do not yet understand how burst accretion is driven \citep{Fischer2022}, we have reached a theoretical and observational consensus that early embedded protostars have more frequent and greater accretion burst events \citep{Vorobyov2015, wspark21}. 
To explore the dynamic processes associated with episodic accretion in protostars, we have to dissect all different kinematical components, from the infalling envelope to the accreting disk, with a high spatial resolution. Decomposing individual kinematic components, however, is not easy during the embedded phase since multiple kinematic features are tangled together along the line of sight. 
Fortunately, we can utilize a variety of molecular transitions to break this degeneracy because specific molecules become abundant under particular physical conditions \citep[e.g.,][]{Oberg2021, Tychoniec2021}.

FU Orionis objects (FUors), which are believed to be in the burst accretion phase in the episodic accretion model, are often associated with nebulae, suggesting that they still have infalling circumstellar envelope material \citep{Takami2018}. In that condition, the infall from envelope to disk can trigger burst accretion through the disk \citep{Vorobyov2005, Vorobyov2015}. Recently, infalling streamers have been revealed by high-resolution ALMA observations of protostars \citep{Pineda2020, Thieme2022, Mercimek2023, jelee23}, and large arm structures in FUors disks have been imaged \citep{Takami2018}. Furthermore, relatively low-velocity shocks with $\sim$3 km s$^{-1}$, which can be produced along the infalling arm structures in the dense inner envelope or at the outer disk boundary, can enhance the abundance of specific molecules, such as SO \citep{Esplugues2014, Sakai2014, Miura2017, vanGelder2021, jelee23}. In addition, the inner disk may be heated above the water sublimation temperature of $\sim$100 K by the burst accretion and can sublimate the ices of Complex Organic Molecules (COMs) together with the water ice \citep{Bianchi2022}. 
To resolve and analyze these physical components associated with infall and accretion, especially the water sublimation radius, FUors in the protostellar phase are likely the best targets \citep{jelee19}.

V883 Ori is a young eruptive star, which is in transition from the Class I to Class II stage. It has a bolometric luminosity of $\sim$220 L$_\odot$ \citep{Furlan2016} at a distance of 388\,pc \citep{jelee19}. The water snow line was resolved in the disk indirectly with the dust continuum intensity distribution \citep[0.1\arcsec]{cieza16} and modeling of the COM emission \citep[0.1\arcsec]{jelee19} and directly with water emission \citep[0.2\arcsec]{Tobin2023}. In addition, V883 Ori is the first disk where COMs, beyond CH$_3$OH and CH$_3$CN, were detected in the midplane \citep{vanthoff18, jelee19}.  In V883 Ori, the envelope still exists \citep{white19}, and the disk is well-developed. In addition, the gaseous disk is still thick and dense enough to be mapped in molecular transitions. 
Previous lower-resolution ($\sim$0.5\arcsec\ to $\sim$1.5\arcsec) ALMA observations \citep[][]{ruiz2017, ruiz2022} have revealed an outflow to the NW (red-shifted, traced by $^{13}$CO) and SE (blue-shifted, traced by $^{12}$CO) as well as the infalling signature of red-shifted absorption against the continuum in HCN and HCO$^+$. 
Despite careful analyses, these resolutions were unable to resolve the detailed kinematics, especially in the boundary between the envelope  and the disk.  

\begin{figure*}[htp]
\centering
{
\includegraphics[width=1.0\textwidth]{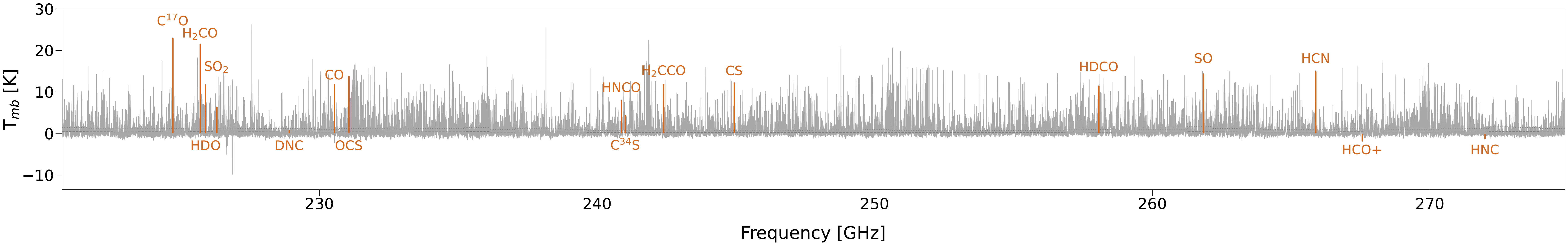}}
\caption{The entire ALMA Band 6 spectra obtained by the spectral scan mode. More than 4500 lines are detected above the three-sigma level. The absorption lines around 226 GHz are due to CN. The orange-colored spectra indicate the molecular lines to be studied in this work, and they are listed in Table \ref{tb:list_simple}.  The spectra were extracted over the COMs emission region ($\sim$0.3\arcsec~in radius), and thus, some molecular lines, such as the HCO$^+$ and HNC lines, which have emission holes inside the water sublimation radius, do not show lines in this plot. %Detailed analyses of the COMs spectra will appear in a separate paper.
\label{fig:all_spec}}
\end{figure*}

\begin{deluxetable*}{cccccc}

\tablecaption{Observation logs\label{tb:obs}}
\tablehead{
\colhead{\bf SGs} & \colhead{\bf Frequency ranges (GHz)}& \colhead{\bf Obs. Date}&\colhead{\bf Baselines} & \colhead{\bf Num. of Antenna} & \colhead{\bf Int. Time ({\bf on-source,} min)}}
\startdata
\multirow{2}{*} {SG1} & \multirow{2}{*}{220.70-- 238.86}  & 2021/05/10 & 15.1 m -- 2.5 km & 44 & 37.9  \\
                   &                                 & 2021/11/09 & 41.4 m -- 3.4 km & 47 & 37.9 \\
\hline
\multirow{3}{*} {SG2} & \multirow{3}{*}{238.70-- 256.86}  & 2021/06/28 & 15.1 m -- 2.0 km & 40 & 22.7  \\
                   &                                 & 2021/06/30 & 15.1 m -- 2.1 km & 42 & 22.7 \\
                   &                                 & 2021/11/11 & 41.4 m -- 3.6 km & 47 & 22.7 \\
\hline
        SG3         & 256.70 -- 274.86           & 2021/11/15 & 41.4 m -- 3.6 km & 52 & 50.6 \\
\enddata
\end{deluxetable*}

In this paper, we introduce the ALMA Spectral Survey of An eruptive Young star, V883 Ori (ASSAY). We analyze the emission distribution of various molecules and the kinematic structures traced by these molecular lines using our new high-resolution Band 6 ALMA observation of V883 Ori. Our spatial resolutions (0.15\arcsec $\sim$ 0.2\arcsec) are high enough to resolve the water sublimation radius, r $\sim$ 0.3\arcsec, in the disk surface, and the maximum recoverable sizes (1.6\arcsec\ to 3.7\arcsec\ equivalent with 600 au to 1500 au) are large enough to cover the infalling inner envelope. Above all, the complete frequency coverage from $\sim$221 to $\sim$275 GHz, with a decent velocity resolution ($\sim$0.6 \kms), allows us to utilize various molecules to explore the different physical/kinematical structures. As a result, our work will complement the investigation by \citet{ruiz2022}.

We describe our observations and data reduction in Section \ref{sec:observation}, and we analyze the spatial distributions of detected molecules in Section \ref{sec:distribution}. The different kinematics traced by specific molecules are explored in Section \ref{sec:kinematics}. In Section \ref{sec:discussion}, we stitch the phenomena found in Sections \ref{sec:distribution} and \ref{sec:kinematics} together to understand how the different physical components are associated with the most recent outbursting episode. Finally, the summary is provided in Section \ref{sec:summary}.

\section{Observation} \label{sec:observation}

The ASSAY (ALMA Spectral Survey of An eruptive Young star, V883 Ori) observations were carried out in band 6 during Cycle 7 (2019.1.00377.S, PI: Jeong-Eun Lee).
The spectral survey covers from $\sim$221 to $\sim$275 GHz with three Science Goals (SGs), which have five tunnings each and four spectral windows (SPWs) per tuning, resulting in 60 SPWs in total. Each SPW covers a bandwidth of 933.5 MHz and a spectral resolution of 488.281 kHz (0.66$\sim$0.53 \kms). The observation logs are listed in Table~\ref{tb:obs}.

The calibration was performed using the CASA pipeline (6.2.1.7) with a quasar, J0423-0120, as the bandpass and flux calibrators and a quasar, J0529-0159, as the phase calibrator. Three phases and one amplitude self-calibrations were carried out for each SG. V883 Ori has emission from many molecular lines, and thus, it is not easy to find line-free channels in the UV plane. Therefore, the spectral line images were cleaned with the continuum, and the line channels were found by requiring either strong extended emission (CO, HCN, HCO$^+$, etc) or, for the COMs emission, a signal higher than the 3 $\sigma$ RMS level in the averaged spectra within an aperture of 0.3\arcsec. Next, the continuum images, cleaned using the line-free channels, were adopted for the self-calibration.  The self-calibration increases the S/N of the continuum images by a factor of 6. The S/N of line data was only slightly improved, however, as the original calibration provided RMS levels close to the theoretical values.

The spectral images were cleaned using the Briggs weighting (robust = 0.5) for SGs 1 and 2 and the Natural weighting for SG 3, to achieve a better resolution. The beam sizes and RMS noise levels for all SPWs are provided in Table~\ref{tb:img} in the Appendix. The continuum was subtracted in the image space, and the continuum level is automatically measured using STATCONT \citep{Sanchez-Monge2018}. To check the consistency across the 60 SPWs, all SPWs were convolved into a beam of 0.25\arcsec $\times$ 0.15\arcsec\ (-77\degree), similar to the poorest resolution of the second SG. 
The averaged continuum intensity within an aperture of 0.3\arcsec\ (red crosses) is plotted in Figure~\ref{fig:cont} in the Appendix. The continuum images using the STATCONT show consistent values within individual SPWs, confirming that the calibrations and the continuum subtraction were well carried out. The second SG shows lower fluxes than the other two SGs but remains within the calibration uncertainty of 5\% \citep{francis20}. 

\section{Spatial Distribution of Molecular Emission} \label{sec:distribution}
Our ALMA observations detect about 4600 lines (Figure~\ref{fig:all_spec}), mostly from COMs. The spectra presented in Figure~\ref{fig:all_spec} have been extracted by the PC1-filtering method developed using the Principle Component Analysis (PCA) applied to the cube data of isolated strong COMs lines \citep{Yun2023}. PCA is one of the multivariate statistics that access the common features veiled in the variation of data \citep{Jolliffe1986}, and recently, PCA has been applied to ALMA observations to assess the distribution of molecular lines in position-position-velocity spaces  \citep[PCA-3D,][]{Oko2021}. Here, PC1 is the first principle component derived from the PCA and represents the most common kinematic and spatial emission structures of COMs. In V883 Ori, COMs emit mainly in the disk with the Keplerian rotation \citep{jelee19, Yun2023}. Therefore, PC1 contains the kinematic information (i.e., Keplerian rotation) that can be used to correct the velocity shifts of spectra.  We apply the PC1-filtering method to the entire frequency range covered by our spectral survey to extract the spectra presented in Figure~\ref{fig:all_spec}. Detailed analyses of the extracted COMs spectra will appear in a separate paper (Jeong et al. submitted). 

In this paper, we focus on simple molecules, which are composed of less than 6 atoms, while selected images of several COMs are also presented. The simple molecules used for this work are marked in Figure~\ref{fig:all_spec} in the orange color and listed in Table~\ref{tb:list_simple}. There are many more lines of simple molecules, detected by our spectral survey but not covered by this work. We use the self-calibrated original cube data, without the beam convolution used in Section \ref{sec:observation} to check for the consistency of all SPWs. We analyze both the 2-D emission distribution at high angular resolution and the kinematics traced by different molecules after identifying detected simple molecules and checking their contamination by other lines based on the spectra presented in Figure~\ref{fig:all_spec}.

The physical parameters of V883 Ori are adopted from the literature as follows. 
The distance of V883 Ori has been updated recently to $388\pm 16$ pc \citep{jelee19}. The systematic velocity of V883 ori is 4.3 \kms\ \citep{cieza16}. The inclination inferred from the ratio between the disk semi-major and semi-minor axes is 38.3 $\pm$ 1 \degree\ \citep{cieza16}, and the position angle of the rotation axis, which is the same as the outflow axis, is $\sim$120\degree \citep{cieza16,ruiz2017}. The blue-shifted outflow is located in the southeast direction from V883 Ori. Thus, the southeast part of the disk is facing toward us. \citet{cieza16} obtained that the disk radius, traced by the C$^{18}$O J=2--1, is 338$\pm$25 au, while the dust disk is limited within a radius of 116 au. Note that the continuum emission extends up to 0.\arcsec8 in our observations as shown in the contours in Figure~\ref{fig:mom_hdo} because of the larger maximum recoverable scale compared to that in \citet{cieza16}. Based on the new distance measurement, the central mass is 1.2 M$_\odot$, estimated using the Keplerian rotation traced by the C$^{18}$O and C$^{17}$O lines \citep{cieza16,jelee19}.\par
\begin{deluxetable*}{ccccc}
\tablecaption{List of simple molecules \label{tb:list_simple}}
\tablehead{
\colhead{\bf No.} & \colhead{\bf molecular transition} & \colhead{\bf Frequency [GHz] (SPWs)} & \colhead{\bf E$_{u}$ [K]}}
\startdata
1 & HDO\tablenotemark{} 3(1,2)-2(2,1) & 225.89672 (SG1 SPW6) & 167.6 \\ 
2 & HDCO 4(3,2)-3(3,1) & 258.07094 (SG3 SPW2) & 102.6 \\
3 & HNCO 11(1,11)-10(1,10) & 240.87573 (SG2 SPW3) & 112.6 \\
4 & OCS 19-18 & 231.06099 (SG1 SPW12) & 110.9 \\ 
5 & C$^{34}$S 5-4 & 241.01609 (SG2 SPW3) & 27.8 \\
6 & SO$_{2}$ 14(3,11)-14(2,12) & 226.30003 (SG1 SPW7) & 119.0 \\
7 & H$_{2}$CO 3(1,2)-2(1,1) & 225.69775 (SG1 SPW6) & 33.4 \\
8 & HCN 3-2\tablenotemark{a} & 265.88643 (SG3 SPW11) & 25.5 \\
9 & HNC 3-2 & 271.98114 (SG3 SPW17) & 26.1 \\
10 & DNC 3-2 & 228.91049 (SG1 SPW10) & 22.0 \\
11 & C$^{17}$O 2-1\tablenotemark{b} & 224.71419 (SG1 SPW5) & 16.2 \\
12 & CO 2-1 & 230.53800 (SG1 SPW11) & 16.6 \\
13 & HCO$^+$ 3-2 & 267.55763 (SG3 SPW12) & 25.7 \\
14 & CS 5-4 & 244.93556 (SG2 SPW7) & 35.3 \\
15 & SO 7(6)-6(5) & 261.84372 (SG3 SPW6) & 47.6 \\
16 & H$_{2}$CCO 12(3,10)-11(3,9) & 242.39845 (SG2 SPW5) & 193.0 \\
\enddata
\tablenotetext{a}{HCN 3-2 has 5 hyperfine components. We refer to \citet{Ahrens2002} for their rest frequencies.}
\tablenotetext{b}{C$^{17}$O 2-1 has 9 hyperfine components. We refer to \citet{tine2000} for their rest frequencies.}
%{C$^{17}$O 2-1 has 6 hyperfine components. We refer to CDMS, Lovas, and JPL databases, which were browsed through Splatalogue (https://splatalogue.online/), for their rest frequencies.}}
\end{deluxetable*}

To investigate the emission distribution and associated kinematics of simple molecules, we utilized various moment maps. 
{\it Moment 0} corresponds to integrated intensity, 
{\it Moment 1} shows the intensity-weighted velocity, 
%{\it Moment 2} shows the intensity weighted velocity dispersion, 
{\it Moment 8} reveals the peak intensity (the maximum value across the channels), and 
{\it Moment 9} represents the peak velocity (the velocity of the peak intensity). 
Note, therefore, that in this paper, the moment 0 and 1 maps follow the standard mathematical definition, however, the moment 8 and 9 have been redefined. Here the moment 8 and 9 maps are defined such that they report similar information as in the moment 0 and 1 maps, respectively. 

We create the moment maps of each molecule by clipping out values below 4$\sigma$. The value of the RMS of each SPW is listed in Table~\ref{tb:img}. Moment 0 and 1 maps are created by integrating the emission over a range of velocity so that there are chances of contamination by other unidentified lines, hyperfine structure lines, or multiple kinematic components. However, moment 8 and 9 maps better present the distribution of the dominating line emission. In this work, we use conventional moment 0 and 1 maps for most lines, but moment 8 and 9 maps are adopted if necessary.

%Figures~\ref{fig:mom_hdo}--\ref{fig:mom3} show the moment 0 and 1 maps of simple molecular transitions. 
The most prominent common feature from the moment 0 maps is the emission hole inside r$\sim$0.1\arcsec. This feature has already been well recognized in the previous ALMA Band 7 observations \citep{jelee19}, and it is caused by the optically thick dust continuum emission, which blocks all the molecular emission \citep{jelee19, ruiz2022}.
The emission size varies significantly between molecules; CO ($>4$\arcsec) and HCO$^+$ ($\sim 3$\arcsec) have the largest and second largest emission regions, respectively, while SO$_2$ ($\sim 0.5$\arcsec) has the smallest emission size. 
%On the other hand, it is hard to measure the CN emission size because absorption dominates over emission for the CN lines. 
Details of the morphological distribution of each molecular emission are presented in the following subsections.
%We here describe the morphological distribution of each molecular emission. 

\subsection{Water sublimation region}

\begin{figure*}[htp]
\centering
\hspace{-1cm}
\includegraphics[width=0.85\textwidth]
{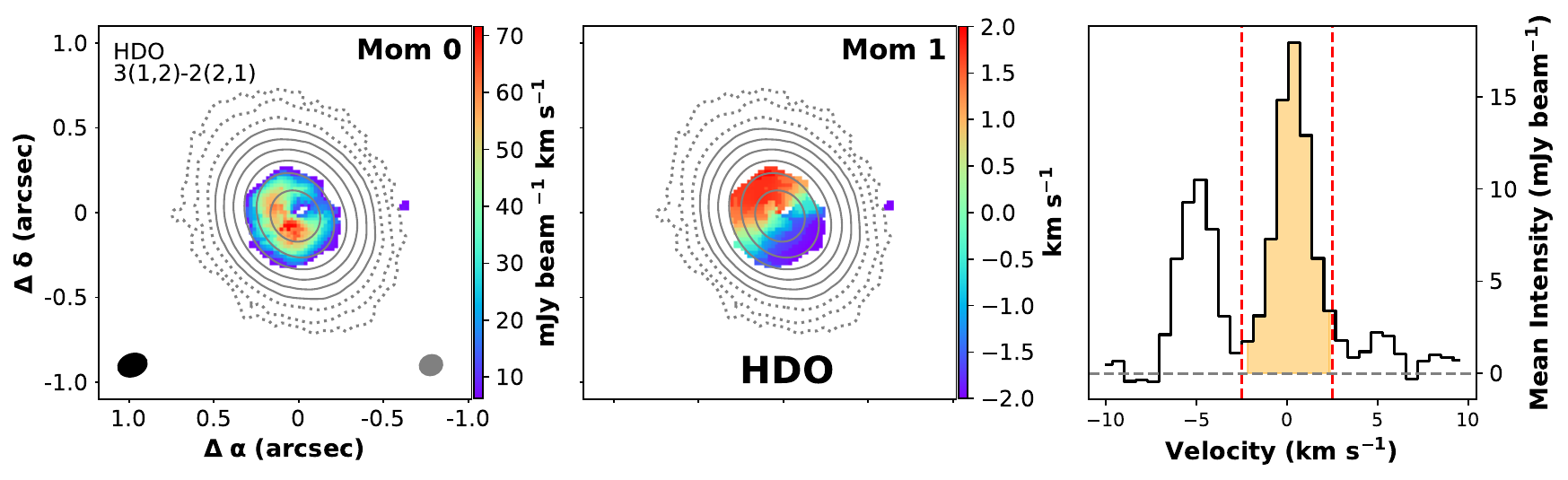}%simple.hdo_hdo.4sig.mom0189.v2.pdf}
\caption{Moment maps and line profile of HDO 3(1,2)-2(2,1), which is the strongest line among 4 HDO lines covered by this spectral survey. Contours in the left (moment 0) and middle (moment 1) present the continuum emission distribution extracted from the SG3 data set, which has the best resolution. The contour levels are 5, 10, 25, 50, 100, 200, 400, 800, and 1600 $\sigma$ ($\sigma$= 0.023 mJy beam$^{-1}$). The first three contours for low intensities are plotted with dotted lines. The boundary of the HDO emission shown in the integrated intensity map (moment 0) is the water sublimation radius {\it in the disk surface}. The disk rotation is clearly seen at the moment 1 map. 
%The velocity dispersion (moment 2) is large toward the center. 
%The peak intensity map (moment 8) and the velocity map at the peak intensity (moment 9) are similar to the moment 0 and 1 maps, respectively, {\bf since there is no contamination by other lines.} 
The ellipses in the lower left and right corners at the moment 0 map represent the beam sizes of the HDO line and continuum, respectively. The contour levels and the beam size of the continuum image are the same in all other figures. The unit of moment 0 and 1 maps are mJy beam$^{-1}$ km s$^{-1}$ and km s$^{-1}$, respectively.
%km s$^{-1}$, and mJy Beam$^{-1}$, respectively. 
The line profile in the right panel is averaged over all pixels with an intensity greater than 4 $\sigma$ after correcting the velocity shift caused by the Keplerian rotation at individual pixels. The velocity range used for the moment maps is marked with the dashed red vertical lines. }
\label{fig:mom_hdo}
\end{figure*}

HDO (Figure~\ref{fig:mom_hdo}): %One highlight from our observational results is {\it the detection of the HDO emission lines in the disk}, directly tracing the water sublimation region. 
V883 Ori is the only disk source where the water snowline is resolved via water emission itself \citep{Tobin2023}.
The moment maps of HDO are presented in Figure~\ref{fig:mom_hdo}; the HDO line adopted in this figure is the same as \citet{Tobin2023} used.
A stronger emission is seen in the southeast half because the southeast part faces us due to the disk inclination.
The line profile in the right panel was extracted over all pixels with intensity above 4 $\sigma$ after correcting the velocity shift caused by the Keplerian rotation at individual pixels. The strong line next to the HDO line is a CH$_3$$^{13}$CHO line.

The boundary of the HDO emission (r$\sim$0.3\arcsec) defines the water sublimation radius {\it in the disk surface}. Within the disk, the water sublimation radius varies with disk height to form an axis-symmetric water sublimation surface. In particular, the water sublimation radius in the disk midplane is referred to as the {\it water snow line}. 
The largest water sublimation radius, as traced by HDO, is about 120 au ($\sim$0.3\arcsec). On the other hand, the water snow line is smaller than 120 au;  \citet{cieza16} and \citet{jelee19} suggested a radial size $\sim$40 au for the water snow line in V883 Ori based on the dust continuum intensity profiles and modeling of the $^{13}$CH$_3$OH line emission distribution, respectively. However, \citet{Tobin2023} argued that the snowline of V883 Ori is $\sim$80 au. A better model is required to constrain the water snow line more accurately.
In this paper, therefore, we refer to the water sublimation radius of 120 au as the boundary of the HDO emission, which must be located on the disk surface.

\subsection{Molecular emission confined within the water sublimation region}

\begin{figure}[htp]
\centering
\includegraphics[width=0.45\textwidth]{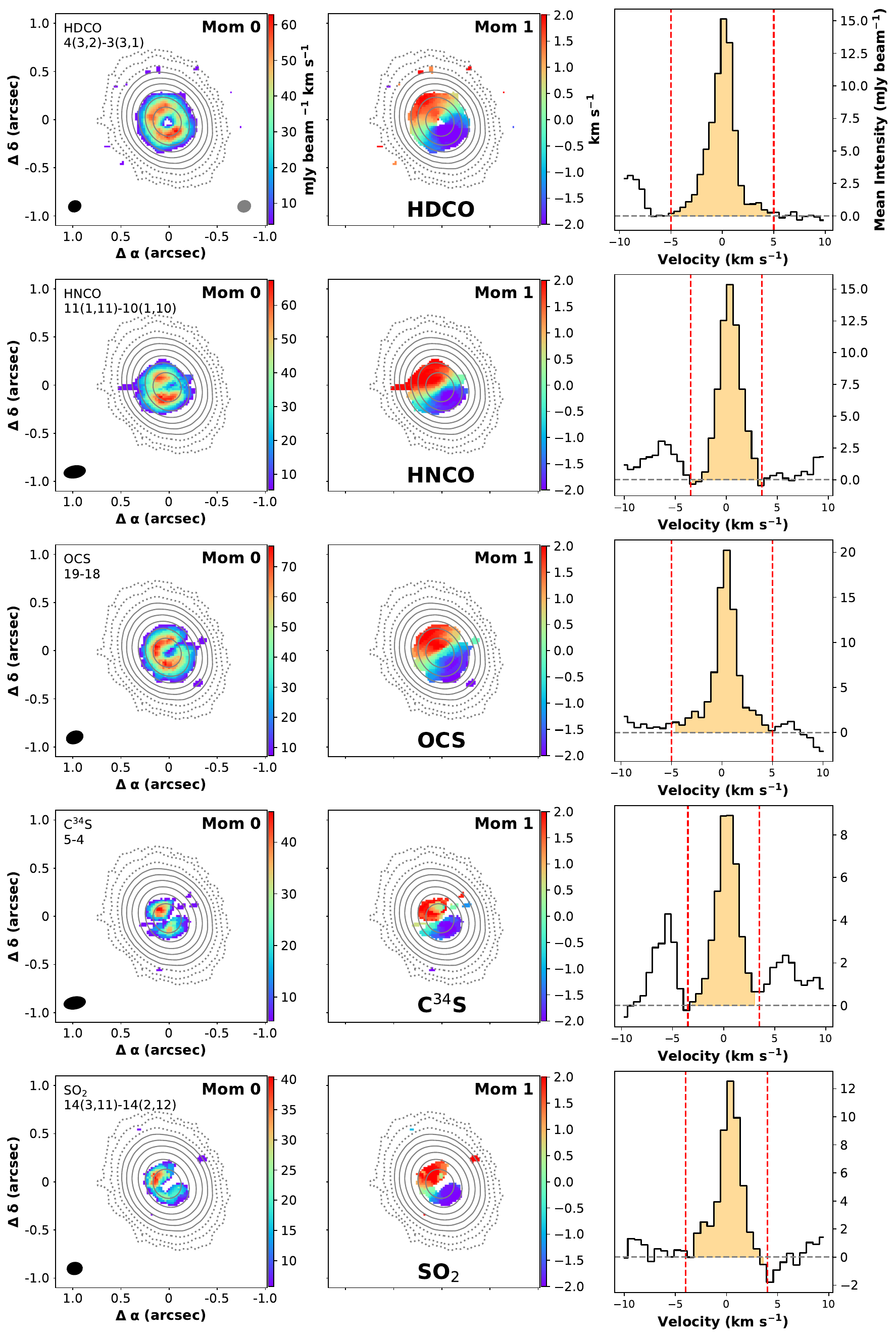}%simple.hnc_c17o.Mom.v3.pdf}
\caption{Moment maps and line profiles of the spectral line showing compact distribution (HDCO, HNCO, OCS, C$^{34}$S, and SO$_2$ lines), whose size is similar to the water sublimation radius.
Contour levels for the continuum are the same as in Figure~\ref{fig:mom_hdo}, and the spectra in the third column are extracted as explained in Figure~\ref{fig:mom_hdo}.
%All emission is well confined within the water sublimation radius traced by the HDO emission.
}
\label{fig:mom3}
\end{figure}

\begin{figure*}[htp]
\centering
\includegraphics[width=0.95\textwidth]{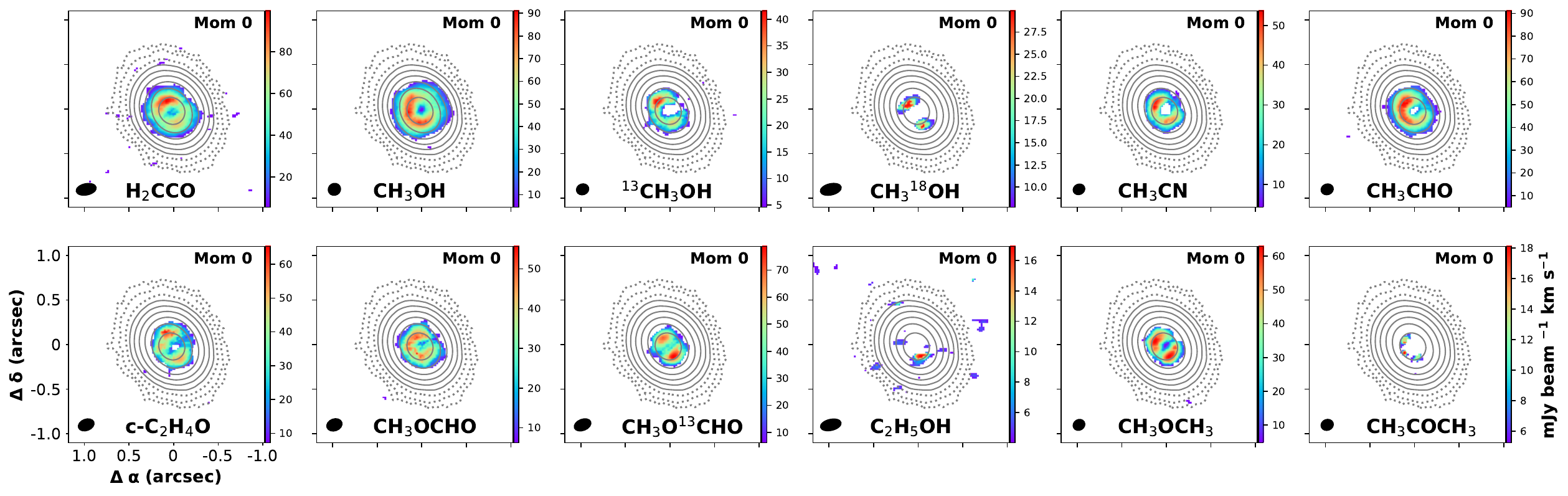}%all_coms.mom0.v4.pdf}
\caption{Integrated intensity (moment 0) maps of selected COMs. The adopted line transition of each COM is listed in Table~\ref{tb:list_coms}}.% All COMs emissions are confined within the water sublimation radius. The detailed analyses of COMs will be presented in a separate work. }
\label{fig:coms}
\end{figure*}

\begin{figure}[htp]
\centering
\includegraphics[width=0.4\textwidth]{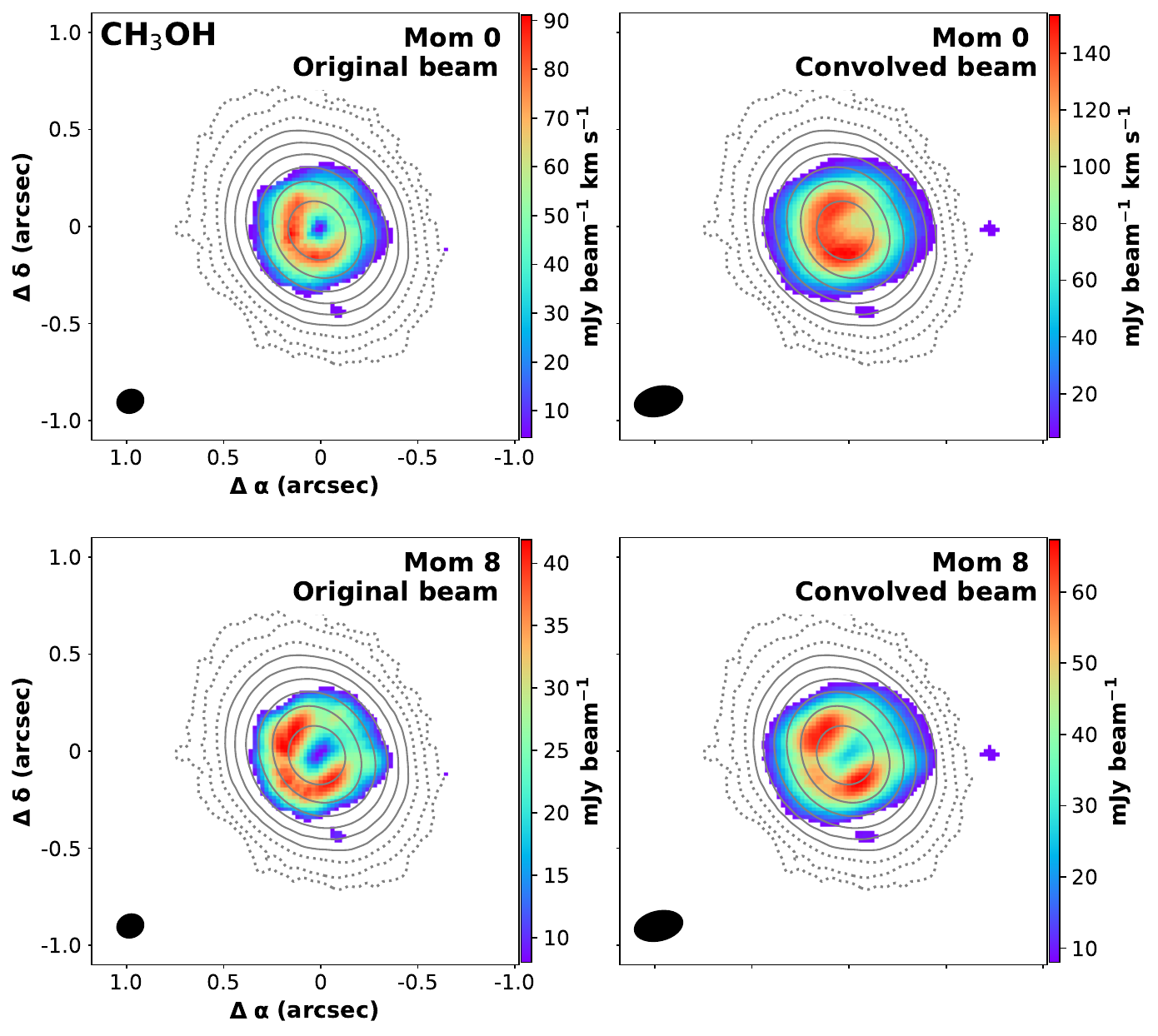} %v883ori.3rd.spw09.org.ch3oh.mom0_8.5sig.subim.v2.pdf}
\caption{Comparison between the original high-resolution images (left) and the images convolved with the elongated beam of the SG2 observations (right). The beam is located in the lower left corner of each image.}
\label{fig:beam}
\end{figure}

%{\bf HDCO, HNCO, OCS, C$^{34}$S, SO$_2$, COMs, CS, and SO (Figures~\ref{fig:mom3}, \ref{fig:coms}, \ref{fig:beam}, and \ref{fig:mom1}):}
HDCO, HNCO, OCS, C$^{34}$S, SO$_2$, and COMs (Figures~\ref{fig:mom3}, \ref{fig:coms}, and \ref{fig:beam}):
The five simple molecules in Figure~\ref{fig:mom3} and COMs in Figure~\ref{fig:coms} trace mainly the water sublimation front and the Keplerian rotation of the inner disk distinctively, as seen in their moment maps. 
%
%The strong CS and SO emission is also confined within the water sublimation radius although they also show extended and faint large emission structure beyond the water sublimation radius as presented in Figure~\ref{fig:mom1}.

Figure~\ref{fig:coms} shows only the integrated intensity (moment 0) maps of 11 selected COM lines, including H$_2$CCO, since moment 1 maps of these lines have similar trends to those of HDO. 
More COMs have been detected in our spectra extracted over the disk (Figure~\ref{fig:all_spec}), but the weak lines of some low-abundance COMs are not strong enough to produce nice 2-D images. 
Over our frequency coverage, COMs have multiple transitions, among which the strongest line in SG3 was selected for their moment maps in each species for the best resolution. If the lines in SG3 are too weak, we used the strongest lines either in SG1 or SG2. Table \ref{tb:list_coms} lists the COMs lines used for the moment 0 maps in Figure~\ref{fig:coms}.

All emissions from COMs, as well as HDCO, HNCO, and OCS, arise exclusively from the central region within the water sublimation radius. Therefore, these molecules could be used as alternative tracers of the water sublimation region in the disk.

%Toward the center of the disk, HDO (Figure~\ref{fig:mom_hdo}) and OCS, as well as SO, H$_2$CO, and H$^{13}$CN (Figure~\ref{fig:mom2} and \ref{fig:mom4}) show broad line widths, while C$^{17}$O, however, shows a very uniform line width. The line widths could be affected by the rotation velocity profile and the emission distribution of molecular lines within the observed beam sizes. The broad line widths might be produced by turbulence probably generated by accretion shock toward the center although a higher angular resolution and higher sensitivity observations at lower frequencies need to study more accurate velocity distributions at the central region.
%HCN shows very broad line widths in both the central region and the ring structure but this is due to the hyperfine {\bf components with separation of -2.3 and +1.8 \kms\ relative to the main component in} the transition. Therefore, we cannot extract real information on the velocity dispersion with this line. 

%\begin{figure*}[htp]
%\centering
%\includegraphics[width=0.6\textwidth]{figure/new_mom/simple.c34s_cn_absor.4sig.mom0189.v4.pdf}%simple.c34s_cn.Mom.v3.pdf}
%\caption{Moment maps of C$^{34}$S, SO$_2$, CN. The contours are the same as in Figure~\ref{fig:mom1}. CN shows some emission beyond the dust disk (the third row). The bottom row shows the moment maps of the CN absorption line below the continuum level, and thus, the integrated intensity (moment 0) and the peak intensity (moment 8) values are all negative. }
%\label{fig:mom5}
%\end{figure*}

C$^{34}$S shows somewhat different emission distributions from the other molecules. 
C$^{34}$S (also HNCO in Figure~\ref{fig:mom3}) has two elongated emission blobs. However, this is not a real emission feature of this molecule, rather the very elongated beam shape obtained by SG2 distorts the emission feature. 

Figure~\ref{fig:beam} shows the comparison between the original CH$_3$OH image and the image convolved with the elongated beam obtained by SG2 to demonstrate the C$^{34}$S image does not show a real emission distribution. The two  disconnected emission blobs seen in HNCO, C$^{34}$S, H$_2$CCO, CH$_3$$^{18}$OH, CH$_3$O$^{13}$CHO are due to the beam shape.

SO$_2$, known as a shock tracer, seems to have stronger emission toward the north than in the south.
Similar characteristics are exhibited by some of the COMs lines, such as $^{13}$CH$_3$OH (see Figure~\ref{fig:coms}).
%This, together with the high-velocity dispersion found for OCS in the northern part of the disk (Figure~\ref{fig:mom4}), 
This may imply that a shock event existed within the inflow motion to the northern part of the disk, as hinted by the HCO$^+$ channel map (see Section \ref{subsec:large_arm_structure}).

%CN shows a strong absorption feature at the map center (the 4th row in Figure~\ref{fig:mom5}).  The line profile of CN (Figure~\ref{fig:spec}) and its moment 1 map for the absorption feature show no velocity structure, indicative of absorption against the continuum peak either by a stationary outer envelope or a close-by interstellar cloud. The CN line also shows a weak emission feature which concentrates at the strong SO$_2$ emission spot in the northern disk, with a hint of extended faint emission beyond the disk (the 3rd column in Figure~\ref{fig:mom5}). 
\vspace*{-0.5cm}
\begin{deluxetable*}{cccc}
\tablecaption{List of selected COMs \label{tb:list_coms}}
\tablehead{
\colhead{\bf No.} & \colhead{\bf molecular transition} & \colhead{\bf Frequency [GHz] (SPWs)} & \colhead{\bf E$_{u}$ [K]}}
%& \colhead{\bf clipping level in moment maps\tablenotemark{a}}}
\startdata
1 & CH$_{3}$OH 2(1,1)-1(0,1) & 261.80574 (SG3 SPW06) & 28.0 \\
%& 4$\times\sigma$$_{SG3,\,SPW06}$ \\
2 & $^{13}$CH$_{3}$OH 9(-1,9)-8(0,8) & 268.63544 (SG3 SPW14) & 107.5 \\
%& 4$\times\sigma$$_{SG3,\,SPW14}$ \\
3 & CH$_{3}$$^{18}$OH 9(-0,9)-8(1,7)E, vt=0 & 250.14877 (SG2 SPW13) & 112.9 \\
%& 4$\times\sigma$$_{SG2,\,SPW13}$ \\
4 & CH$_{3}$CN 14(4)-13(4) & 257.44813 (SG3 SPW1) & 207.0 \\
%& 4$\times\sigma$$_{SG3,\,SPW1}$ \\
5 & CH$_{3}$CHO 14(7,8)-13(7,7)E, vt=0 & 269.83409 (SG3 SPW15) & 207.5 \\
%& 4$\times\sigma$$_{SG3,\,SPW15}$ \\
6 & c-C$_{2}$H$_{4}$O 8(1,8)-7(0,7) & 235.10502 (SG1 SPW16) & 52.4 \\
%& 4$\times\sigma$$_{SG1,\,SPW16}$ \\
7 & CH$_{3}$OCHO 19(3,17)-18(3,16)E &  225.60882 (SG1 SPW06) & 116.7 \\
%& 4$\times\sigma$$_{SG1,\,SPW06}$ \\
8 & CH$_{3}$O$^{13}$CHO 22(1,22)-21(0,21), v=1-1 & 236.27864 (SG1 SPW18) & 132.8 \\
%& 4$\times\sigma$$_{SG1,\,SPW18}$ \\
9 & C$_{2}$H$_{5}$OH 15(1,15)-14(0,14), anti & 251.56647 (SG2 SPW15) & 97.6 \\
%& 3$\times\sigma$$_{SG2,\,SPW15}$ \\
10 & CH$_{3}$OCH$_{3}$ 24(5,20)-24(4,21)EE & 260.00439 (SG3 SPW04) & 308.9 \\
%& 4$\times\sigma$$_{SG3,\,SPW04}$ \\
11 & CH$_{3}$COCH$_{3}$ 26(1,25)-25(2,24)EA & 268.28371 (SG3 SPW13) & 181.7 
%& 4$\times\sigma$$_{SG3,\,SPW13}$
\enddata
%\tablenotetext{a}{We create the moment maps of each molecule by clipping out values below this level. The value of the RMS of each SPW is listed in Table~\ref{tb:img}.}
\end{deluxetable*}

\subsection{Inner and outer disks}

\begin{figure}[htp]
\centering
\includegraphics[width=0.45\textwidth]{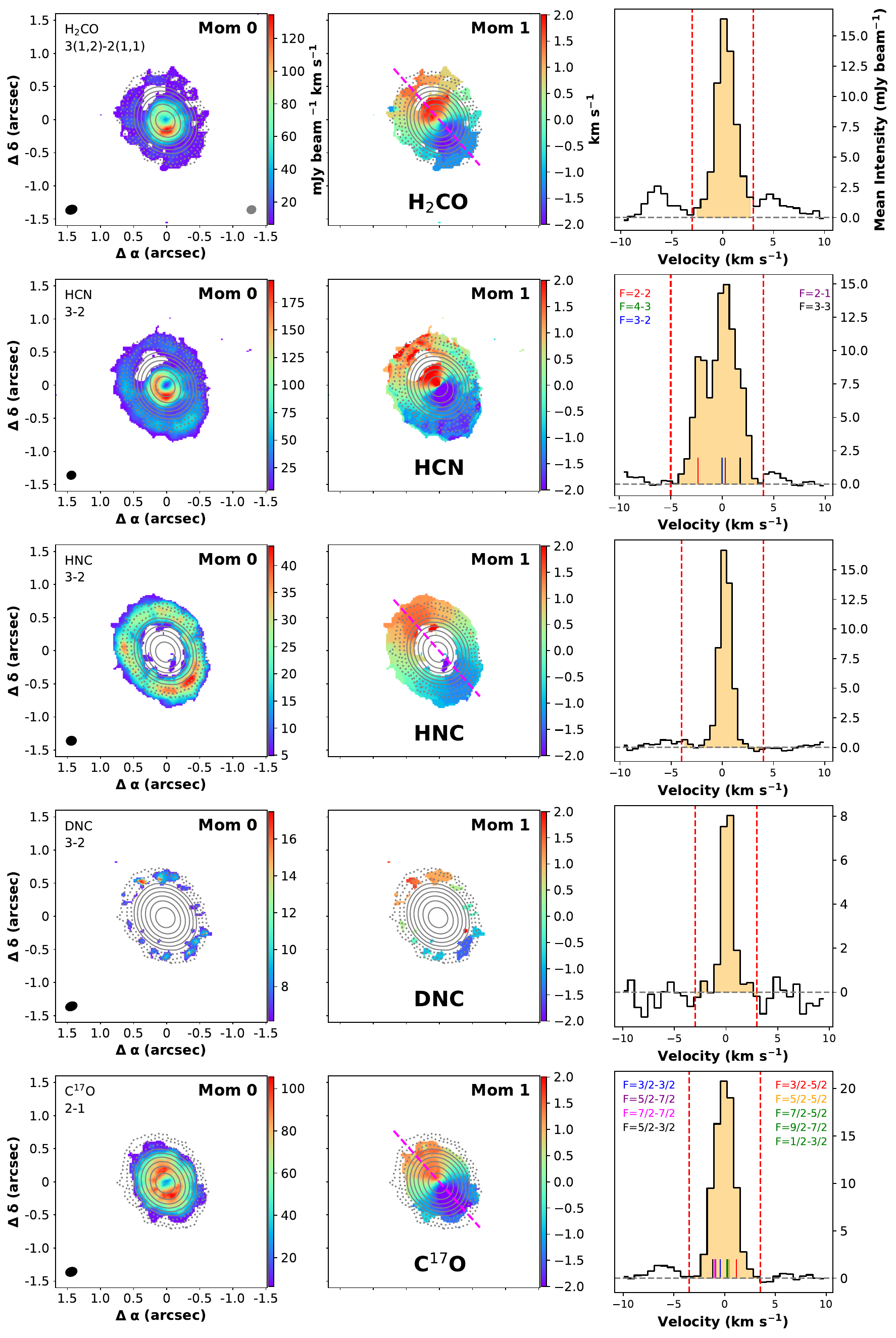}%simple.h2co_hc3n.4sig.mom0189.v3.pdf}
\caption{Moment maps and line profiles of molecules (H$_2$CO, HCN, HNC, DNC, and C$^{17}$O) showing emission in the outer disk.
Contour levels for the continuum are the same as in Figure~\ref{fig:mom_hdo}.
%The moment 2 map of HCN indicates a high-velocity dispersion compared with the other lines. 
Moment 1 map of HCN is affected by its hyperfine structure, whose components are marked in the HCN averaged line profile with vertical solid lines. C$^{17}$O J=2--1 also has a hyperfine structure, as marked in its averaged line profile, but they are not resolved with our velocity resolution.
%The moment 1 map of C$^{17}$O shows the best disk rotation since the C$^{17}$O emission does not have any prominent substructures. 
The magenta dashed lines at the moment 1 maps of H$_2$CO, HNC and C$^{17}$O indicate the cut used for the PV diagram in Figure~\ref{fig:pv_dia}.
The spectra in the third column are extracted as explained in Figure~\ref{fig:mom_hdo}.}
\label{fig:mom2}
\end{figure}

H$_2$CO and HCN (Figure~\ref{fig:mom2}):
These molecules trace the largest gas disk at the size of r$\sim$0.8\arcsec. Disk emission traced by these molecules appears to be divided into two regions: one confined at the center within the water sublimation radius of 0.3\arcsec\ and the other confined within a ring structure between $\sim$0.5\arcsec\ and 0.8\arcsec, where the dust continuum intensity still exists. 
Although the emission seems confined within the disk, these two molecular lines also trace the infalling envelope, as seen in the redshifted absorption against the continuum (see Section \ref{sec:infall}). 

The HCN line has hyperfine structures as marked in its line profile (the third column of the second row in Figure~\ref{fig:mom2}). These hyperfine components contaminate the moment 1 map of HCN 3-2. Therefore, the moment 9 map better presents the velocity distribution of the peak intensity for the main component.
The HCN map also shows arm-like structures connecting the outer disk and the inner disk (See section~\ref{subsec:large_arm_structure}  for the detailed discussion).
%These spiral-like structures connect the inner disk and the outer disk. Their emission and velocity structures seem to connect to the large HCO$^+$ arm-like structures stretched beyond the outer boundary of the disk, as presented in the HCO$^+$ channel map (Figure~\ref{fig:chanmap}). 

%\subsection{Innter and outer disks}

%{\bf D$_2$CO, H$^{13}$CN, HC$^{15}$N, and HC$_3$N (Figure~\ref{fig:}):}
%D$_2$CO, H$^{13}$CN, and HC$^{15}$N show two distinct emission structures with a gap between the two structures, while H$_2$CO has a faint outer part, which is not clearly separated from the strong central emission. 

\subsection{A ring structure in the outer disk}
HNC and DNC (Figure~\ref{fig:mom2}): HNC and DNC trace only the ring structure at the outer disk. The ring is the most prominent in HNC. 
The size of the emission hole of HNC is found to be similar to the water sublimation radius. \citet{Fraser2001} reported that water would sublimate at temperatures greater than $\sim$100 K for interstellar conditions. At the emission hole, HNC could therefore be converted to HCN via
%The emission hole of HNC is well consistent with the water sublimation radius, inside which the temperature is greater than $\sim$100 K \citep{Fraser2001}. Inside the water sublimation radius, HNC is converted to HCN via 
the reaction ${\rm HNC + H} \rightarrow {\rm HCN + H}$ \citep[with an energy barrier lower than 200 K,][]{Hirota1998,Graninger2014,mhJin2015,Hacar2020} because this reaction is efficient above 40~K. A chemical time scale also depends on the hydrogen number density and, thus, could be much shorter in the inner disk than in the outer disk, resulting in a clear HNC ring structure. More detailed chemical models are needed to investigate this conversion \citep{Long2021}.  %Because of this chemical conversion between HCN and HNC, the HNC/HCN ratio has been considered a good gas temperature probe in protoplanetary disks \citep{Long2021}.} 

Figure~\ref{fig:hcnazimuthal} shows the intensity distribution comparison between the HCN and HNC lines; the HNC emission is missing within the hot inner disk, while the HCN emission shows two peaks, at 0.2\arcsec\ and 0.7\arcsec. 
The velocity distribution of the ring mainly follows the disk rotation well (see Section \ref{sec:kinematics}).

\subsection{A tracer of the dust disk}
\begin{figure}[htp]
\centering
\begin{tabular}{c}
\includegraphics[width=0.47\textwidth]{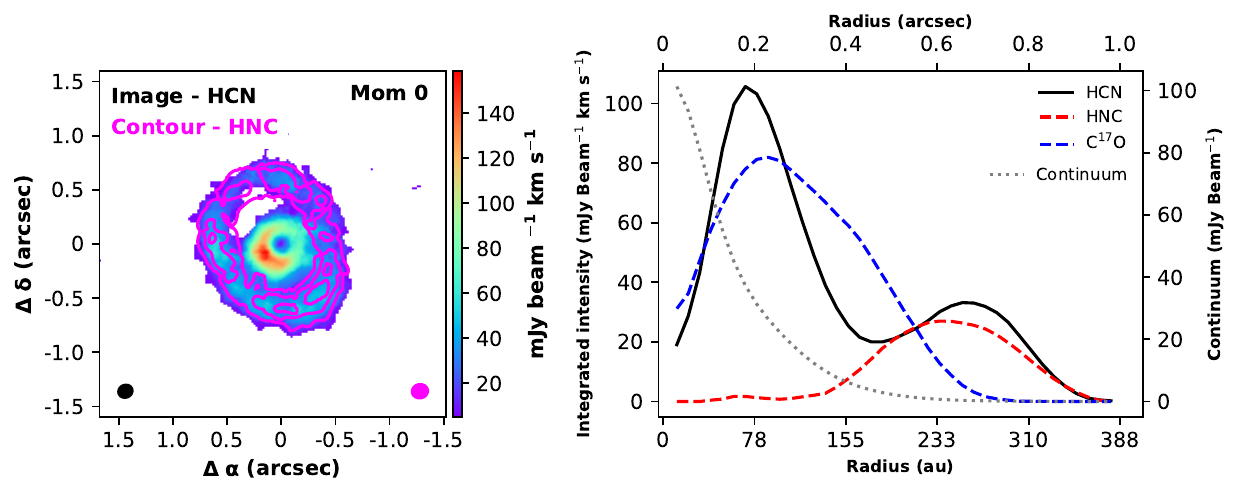}\\%hcop_hdo.hcn_hnc.cont.R_I.v4.pdf}
\end{tabular}
\caption{Integrated intensity distribution comparison between the HCN and HNC lines. In the left panels, the magenta contours show the moment 0 map of the HNC line on top of the moment 0 image of HCN. Contour levels for the molecule are 15, 25, and 35 mJy beam$^{-1}$ km s$^{-1}$. The ellipse in the lower left corner represents the beam size of images, while the ellipse in the lower right corner shows the beam size of contours.
The right panels present the azimuthally averaged intensity profiles of HCN (black solid line) and HNC (red dashed line) as well as the continuum (dotted grey line). 
%Intensity profiles are scaled up and down to match the peak values for better comparisons. 
%Anti-correlation between HDO and HCO$^+$ (HCN and HNC) is well presented. In the right, 
The blue dashed line shows the C$^{17}$O line intensity profile for comparison.}
\label{fig:hcnazimuthal}
\end{figure}

C$^{17}$O (bottom panels of Figure~\ref{fig:mom2}):
In contrast to the other presented lines, except for the emission hole inside 0.1\arcsec, the C$^{17}$O J=2-1 emission distribution (moment 0) is most consistent with the dust emission without any substructure. 
Its outer boundary is close to the peak of the HNC emission (see the blue dashed line in Figure~\ref{fig:hcnazimuthal}).
%its size is similar to the inner boundary of the HNC and DNC emission. 
The high dust opacity causes the missing emission at the center.
In addition, the Keplerian rotation is well recognized (see its moment 1 map in Figure~\ref{fig:mom2}), as seen in the C$^{17}$O J=3-2 emission presented by \citet{jelee19} (see Section \ref{sec:disk_rot}). 
C$^{17}$O is known to be  a good tracer of disk \citep{vanHoff2020} since its lines are optically thin. In addition, the temperature of the V883 Ori disk must be greater than the CO sublimation temperature ($\sim$20 K) in most regions of the disk because of its high central luminosity and viscosity heating in the disk midplane. 
%The C$^{17}$O J=2-1 line has hyperfine structures, which are not resolved in our spectrum.

%\begin{figure*}[htp]
%\centering
%\includegraphics[width=0.6\textwidth]{figure/new_mom/simple.ch3oh_ocs.4sig.mom0189.v4.pdf}%simple.ch3oh_ocs.Mom.v3.pdf}
%\caption{Moment maps of CH$_3$OH, HDCO, D$_2$CO, HNCO, and OCS. The contours are the same as in Figure~\ref{fig:mom1}. Except for D$_2$CO, all emission is confined within the water sublimation radius traced by the HDO emission. D$_2$CO shows some emission beyond the dust disk, probably along the ring structure that appears in the H$_2$CO emission map.}
%\label{fig:mom4}
%\end{figure*}

\subsection{Large-scale structures in the envelope}

\begin{figure}[htp]
\centering
\includegraphics[width=0.45\textwidth]{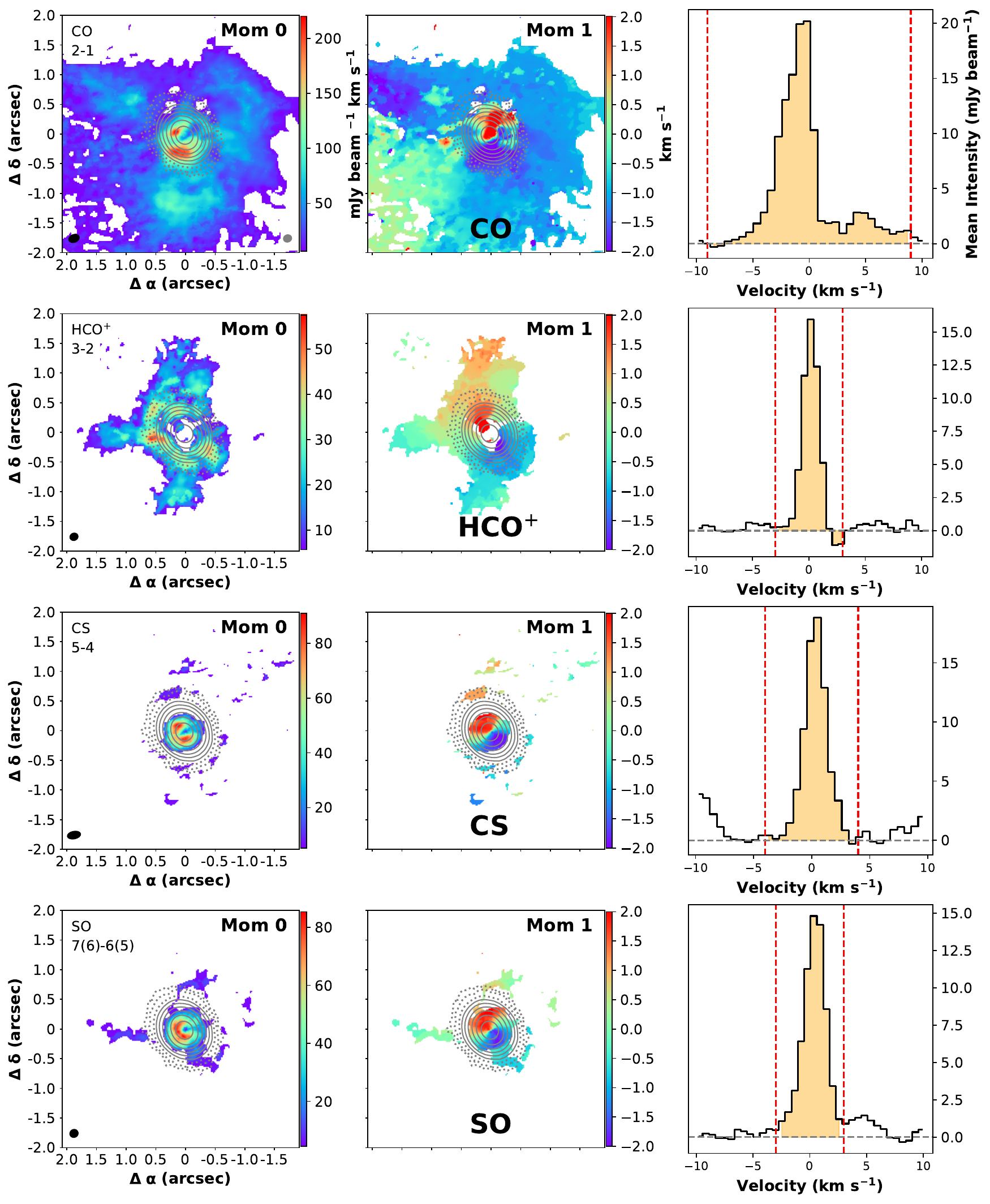}%simple.co_so.4sig.mom0189.v2.pdf}
\caption{Moment maps and line profiles of molecules (CO, HCO$^+$, CS,  and SO) having extended-envelope components in the maps. Contour levels for the continuum are the same as in Figure~\ref{fig:mom_hdo}, and the spectra in the third column are extracted as explained in Figure~\ref{fig:mom_hdo}. 
%CS and SO show the disk rotation within the water sublimation radius.
}
\label{fig:mom1}
\end{figure}

\begin{figure}[htp]
\centering
\includegraphics[width=0.5\textwidth]{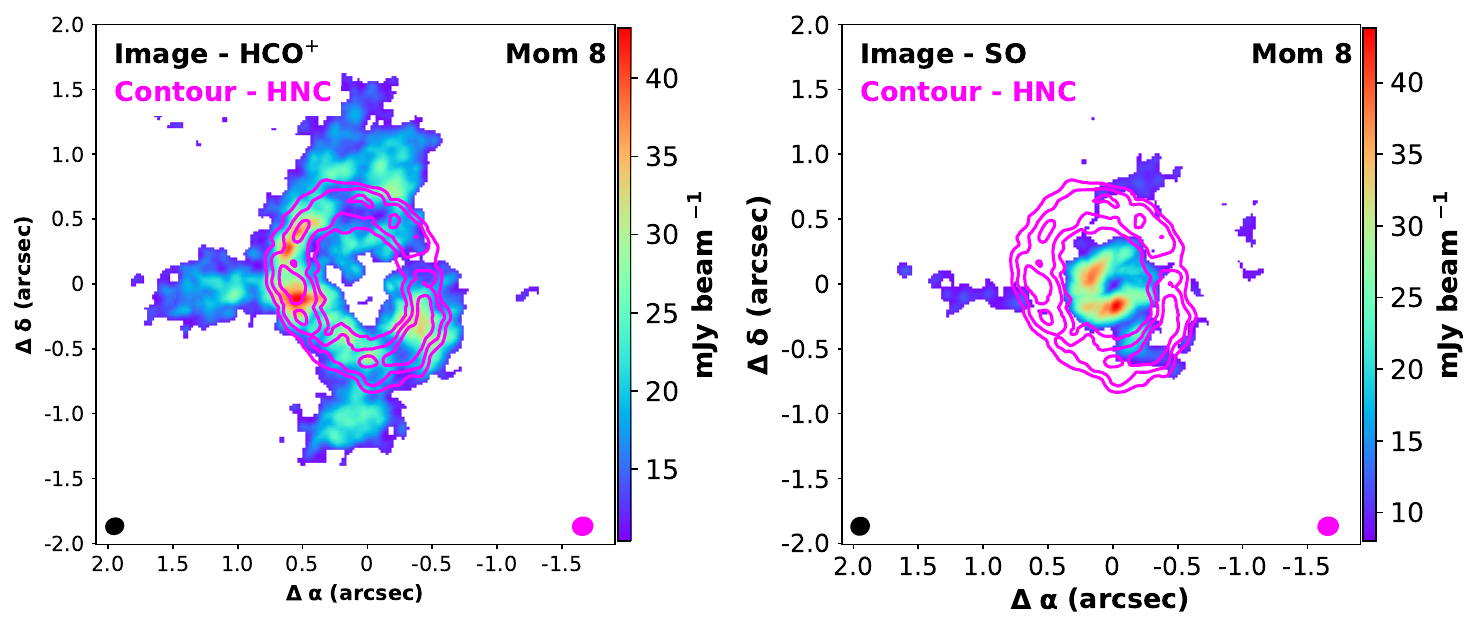}
\caption{Comparisons between HNC and HCO$^+$ (left) or HNC and SO (right).
%The HCO$^+$ emission is the strongest along the HNC ring, while the SO emission is concentrated in the inner disk with arm-like faint structures stretched out beyond the ring. 
The ellipse in the lower left corner represents the beam size of images, while the ellipse in the lower right corner shows the beam size of contours. The contour levels are 14.2, 21.2, and 29.2 mJy beam$^{-1}$.}
\label{fig:ring}
\end{figure}

CO, HCO$^+$, CS, and SO (Figure~\ref{fig:mom1}):  The emission of these molecules appear at scales larger than 1\arcsec, and thus, likely trace the envelope material of V883 Ori, whose morphology is far from spherically symmetric.  First, the CO J=2--1 map shows a clumpy and complex emission distribution. The moment 1 map shows a similar velocity feature to the map of the same CO line reported by \citet{ruiz2017}; the CO emission traces the blue-shifted outflow cavity walls with a wide opening angle. 
The HCO$^+$ emission is strongest along a ring-like structure, which is consistent with the HNC emission distribution (left panel of Figure~\ref{fig:ring}).

Within the central region of $\sim$0.3\arcsec, the HCO$^+$ emission is missing, probably a combination of its destruction by H$_2$O \citep{Leemker2021} and the dust optical depths effects \citep{ruiz2022}. 
Strong CS and SO emission is also confined within the water sublimation radius, although extended and faint large emission structures are seen beyond the water sublimation radius. These molecular lines show the disk rotation within the water sublimation radius.

\begin{figure}[htp]
\centering
\begin{tabular}{c}
\includegraphics[width=0.47\textwidth]{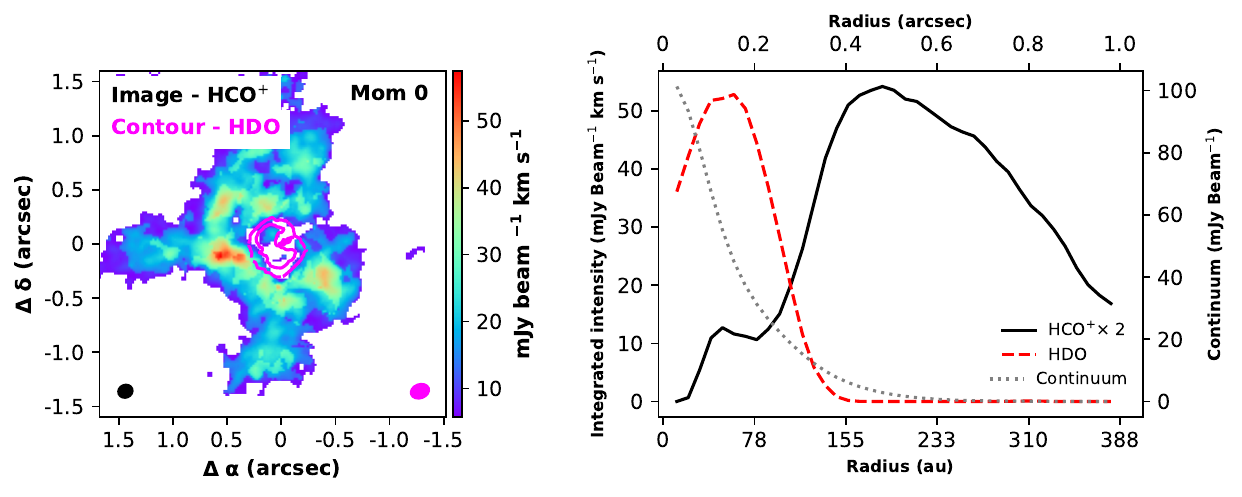}%hcop_hdo.hcn_hnc.cont.R_I.v4.pdf}
\end{tabular}
\caption{Integrated intensity distribution comparison between the the HCO$^+$ and HDO lines. In the left panels, the magenta contours show the moment 0 maps of the HDO line on top of the HCO$^+$ moment 0 image. Contour levels are 15 and 45 mJy beam$^{-1}$ km s$^{-1}$. The ellipse in the lower left corner represents the beam size of images, while the ellipse in the lower right corner shows the beam size of contours. The right panels present the azimuthally averaged intensity profiles of HCO$^+$ (black dotted line) and HDO (red dashed line) as well as the continuum (dotted grey line). Intensity profiles are scaled up and down to match the peak values for better comparisons. 
%Anti-correlation between HDO and HCO$^+$ (HCN and HNC) is well presented. 
 }
\label{fig:hcopazimuthal}
\end{figure}

Figure~\ref{fig:hcopazimuthal} compares the integrated intensity distributions of the HCO$^+$ and HDO lines; the HDO emission lies well inside the HCO$^+$ emission hole, indicative of a chemical anti-correlation. 
The small peak of the HCO$^+$ intensity profile around 0.1\arcsec\ in the right panel (solid black line) may be caused by the contamination of a CH$_3$OCHO line in the vicinity of the HCO$^+$ line although the averaged HCO$^+$ line profile over the whole cube image does not show a blended feature in Figure~\ref{fig:mom1}. This contamination can happen only inside the water sublimation radius since all COMs are frozen on grain surfaces beyond the water sublimation radius.

%Several large arm-like features stretch outward from the HCO$^+$ ring structure. 
HCO$^+$ emission shows several large arm-like structures extending from the outer disk. %the emission ring.
SO also shows faint arms (right panel of Figure~\ref{fig:ring})  with a length of $\sim$1\arcsec, but the strong emission is confined only within the emission hole of the HCO$^+$. Therefore, the strong SO emission distribution coincides with the HDO emission.
The overlayed HNC emission distribution on the SO image 
clearly demonstrates the SO emission is confined within the water sublimation radius with three faint large arm structures.
The SO arms stretch to the east and the northwest along corresponding HCO$^+$ structures. The southern SO arm does not show a clear counterpart in HCO$^+$. 
CS shows the same compact emission in the disk as SO, while the faint extended structures traced by CS do not coincide with those in SO. 

These arm-like structures are in Figure~\ref{fig:ring}, possibly streamers infalling from the envelope to the disk. 
Recently, narrow structures that asymmetrically feed gas from the envelope to the scale of the circumstellar disks \citep[]{Pineda2020, Bianchi2022, Thieme2022, jelee23} have been detected in various molecular tracers and called infalling streamers. According to other observations \citep[]{Valdivia-Mena2022, jelee23}, SO is likely a good tracer of infalling streamers from the inner envelope to the disk. However, the velocity resolution of our data is not high enough to study the kinematics of the arm-like structures in detail. To confirm whether the SO weak emission structures are tracing the true infalling streamers, we need a better spectral resolution as well as a better sensitivity.

\begin{figure*}[htp]
\centering
\includegraphics[width=0.9\textwidth]{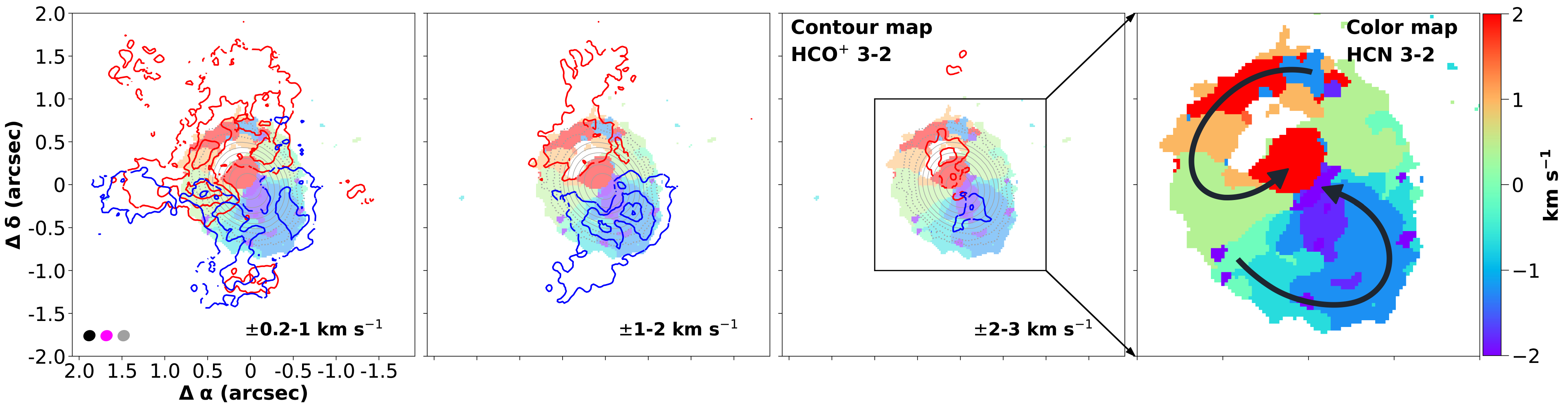}
\caption{ The channel maps of HCO$^+$ (contours) on top of the moment 9 map of HCN (colors). Red and blue contours in each panel are the HCO$^{+}$ intensity integrated over the velocity ranges denoted in the lower right. 
Contour levels are from 5.7 to 57.5 mJy beam$^{-1}$ km s$^{-1}$ in step of 13.0 mJy beam$^{-1}$ km s$^{-1}$. The ellipses in the lower left represent the beam size of the HCN (black), HCO$^{+}$ (magenta), and continuum (gray), respectively. 
%The spiral structures of HCN in the outer disk are connected with the large arm-like structures of HCO$^+$. 
The rightmost panel presents the zoom-in moment 9 map of HCN with bolder colors.)
}
\label{fig:chanmap}
\end{figure*}

\section{Kinematics and Dynamics} \label{sec:kinematics}

\subsection{Large arm-like and small spiral-like structures}\label{subsec:large_arm_structure}
As indicated above, HCO$^+$ and SO show $\sim$1\arcsec\ scale of arm-like structures connected to the disk (see Figure~\ref{fig:ring}). 
%The SO narrow arm-like structures are very faint compared to the central emission inside the disk. 
The velocity distributions (see moment 1 maps in Figure~\ref{fig:mom1}) along the three arms of HCO$^+$ and SO show velocity gradients toward the disk. 
The HCO$^+$ channel maps are presented as contours in Figure~\ref{fig:chanmap}; the outside-in flows are detected at both red- and blue-shifted velocities.
%CO also shows a velocity profile of flows toward the center, counterclockwise both in the northern and southern hemispheres within the dust continuum disk (see the central part of its moment 1 map).
Several high-resolution images by ALMA show relatively large-scale arm structures resembling those detected here \citep[e.g., ][for IRAS 03292+3039, IRAS 04239+2436, respectively]{Pineda2020, jelee23}.

\begin{figure*}[htp]
\centering
\hspace{-1cm}
\includegraphics[width=1.04\textwidth]{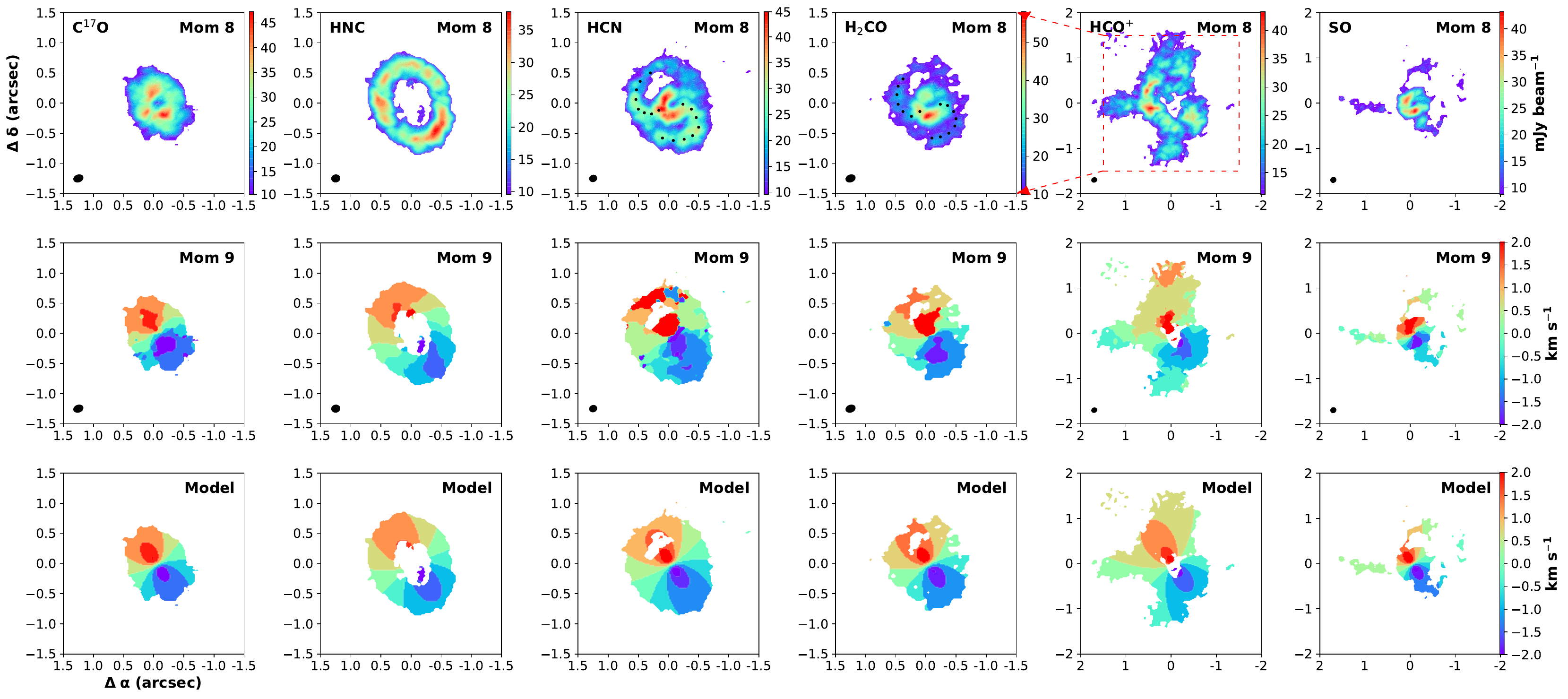}
\caption{The observed moment 8 and 9 maps and the modeled moment 9 maps. The model is based on  Keplerian rotation around a 1.2 \msun\ source with an inclination of 38\degree\ and a PA of 120\degree.
%The peak intensity of HCN (also H$_2$CO) distributes along the outer ring-like region, and spiral-like structures seem to exist in the south and north of the outer disk along the black dots. 
The black dots are plotted at the moment 8 maps of HCN and H$_2$CO to guide along the spiral-like structures.}
\label{fig:model_disk}
\end{figure*}

There are also hints of spiral-like structures connecting the outer and inner disks in H$_2$CO and HCN. 
The peak velocity map of HCN 3-2 (see the rightmost panel in Figure~\ref{fig:chanmap} for a zoom-in image) shows outside-in flow-like structures both in the blue- and red-shifted velocities, as guided by the arrows.
Like HCN, the peak intensities of H$_2$CO also distribute along the outer ring-like region, and spiral-like structures seem to exist in the south and north of the outer disk along the black dots in Figure~\ref{fig:model_disk}.
These types of spiral arms have been resolved within dust disks \citep{cflee2020}. Such spiral structure within the disk can be mainly induced by the gravitational instability of the disk or interaction with a companion \citep{cflee2020}. 

The comparison between the observed and modeled moment 9 maps of C$^{17}$O, HCN, H$_2$CO, and HCO$^+$ show some difference (Figure~\ref{fig:model_disk}). 
The model velocity map was first generated using the disk parameters (i.e., inclination and position angle) and the central mass (see Section \ref{sec:distribution}). Then, it was convolved with the observational spatial and spectral resolutions and masked by the emission region of each molecule for comparison. 
The C$^{17}$O moment 9 map, which is the best tracer of the Keplerian disk, is consistent with the model. In the moment 9 maps of other molecular lines, some differences are notable. It is seen especially in the outer disk in HCN, in the inner disk in H$_2$CO, and along the arm-like structure in HCO$^+$. %However, some differences are notable, especially in the outer disk in HCN, in the inner disk in H$_2$CO, and along the arm-like structure in HCO$^+$. 

For a simple examination of the spiral structures, we follow the analysis presented by \citet{cflee2020} to fit the peak intensity positions of HCN along the spiral arms in the polar coordinates with two spiral shapes: the logarithmic spiral with $R=R_0 e^{a\theta}$, developed by the gravitational instability and the Archimedean spiral with $R=R_0 + b\theta$, induced by the interactions with a companion. 
\begin{figure}[htp]
\centering
\hspace{5mm}\includegraphics[width=0.43\textwidth]{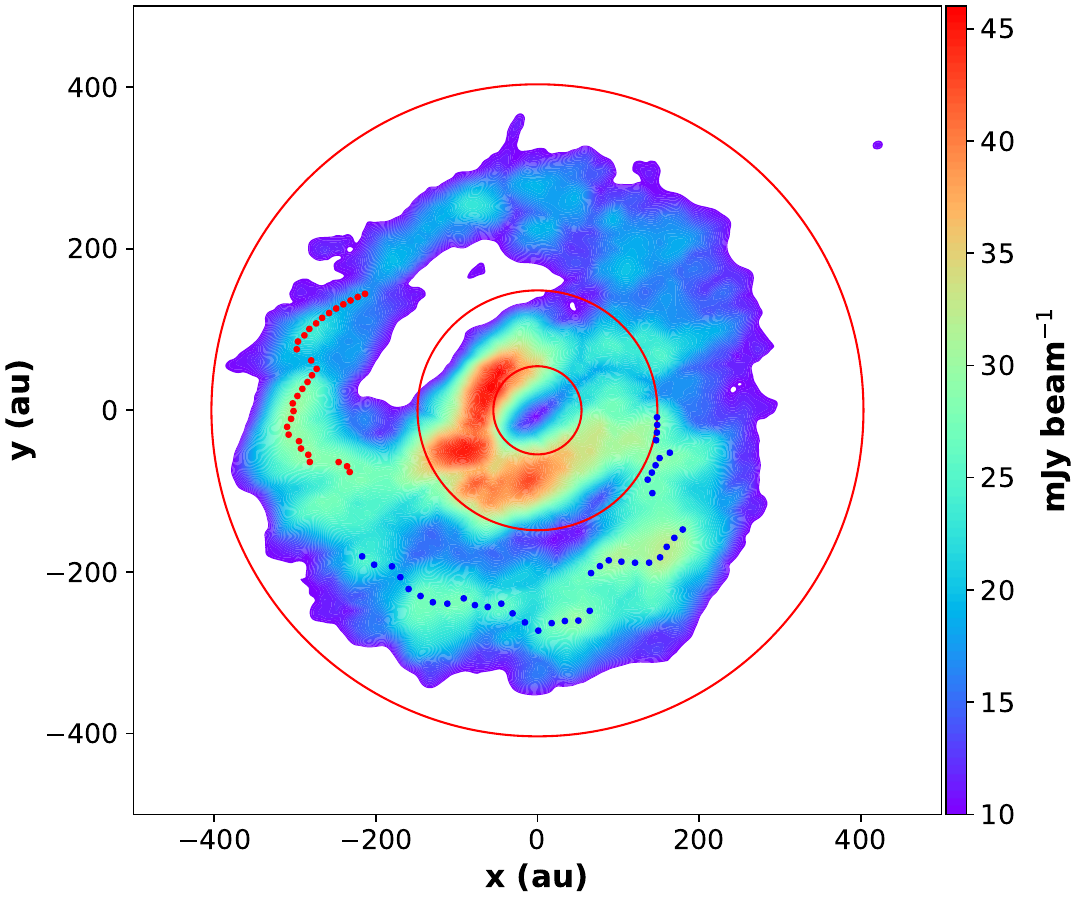}
\includegraphics[width=0.40\textwidth]{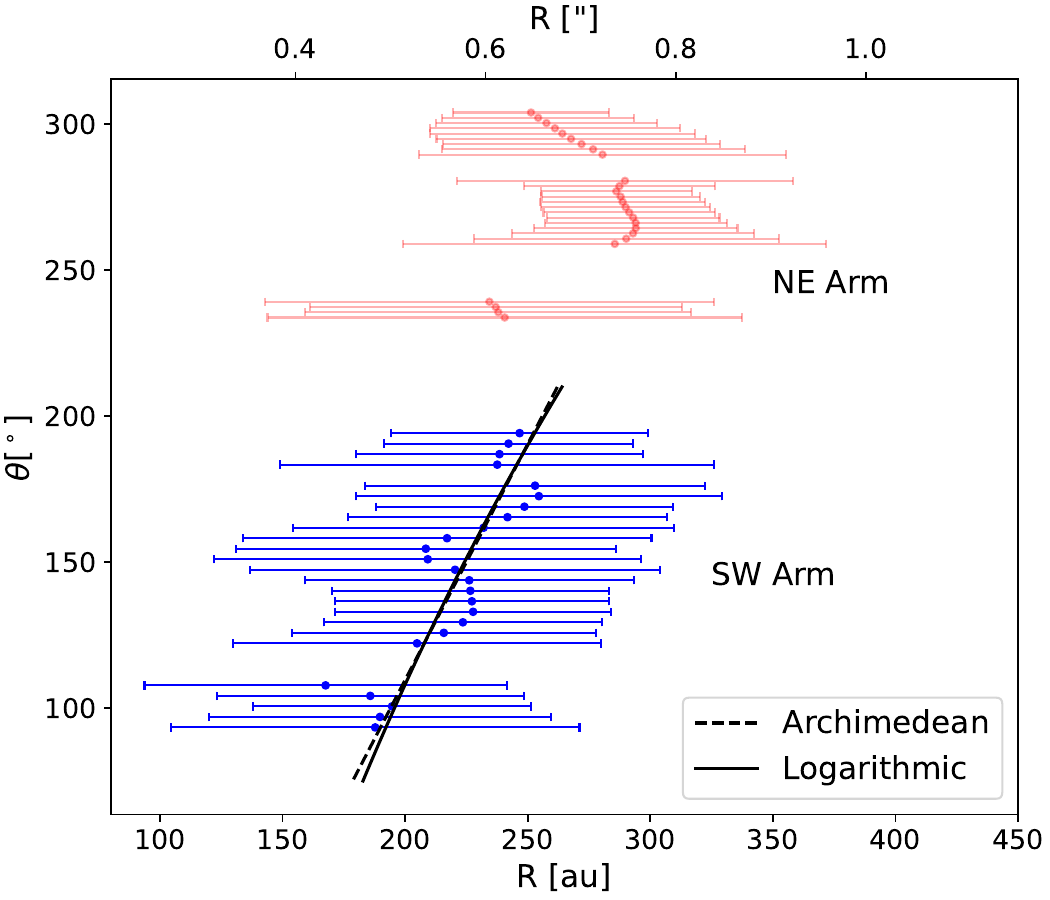}
\caption{Deprojected moment 8 map of HCN and the peak positions of the spiral arms (top) and the spiral profiles of the two arms (bottom) using the inclination of 30 degrees. Three red circles in the deprojected moment 8 map denote the radii of $\ln\, r$ (au) = 4, 5, and 6, respectively. Only the SW arm (blue) has a spiral-like profile and can be fitted with an Archimedean or a logarithmic spiral.}
\label{fig:r_theta}
\end{figure}
Figure~\ref{fig:r_theta} shows fits for these spiral profiles to the two arms of HCN. The northern arm does not reveal a spiral profile, while the southern arm is well-fit. It is unclear, however, which spiral shape fits better the spiral arm.

\begin{figure}[htp]
\centering
\includegraphics[width=0.4\textwidth]{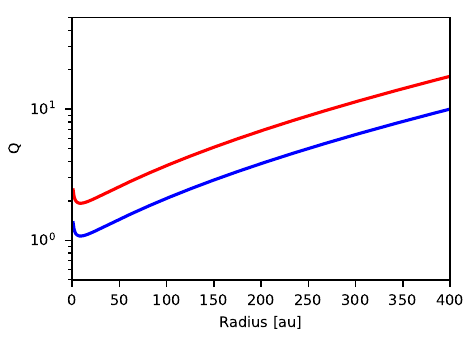}
\caption{Toomre Q parameter for the V883 Ori disk. The red and blue lines indicate the Q parameters using the temperature distribution, $T(r)=100~{\rm K} \times \sqrt{r_{\rm snow}/r}$, in the burst and the quiescent phase.} %The water snow line ($r_{\rm snow}$) in the burst and quiescent phases are located at 39 au and 4 au, respectively.}
\label{fig:q_parameter}
\end{figure}

The gravitational instability in the disk can be quantified by the Q parameter \citep[e.g.,][]{Toomre1964,Kratter2016}:
\begin{equation}
    Q = \frac{c_s \Omega}{\pi G \Sigma},
\end{equation}
where $c_s$ is the sound speed, $\Omega$ is the Keplerian angular velocity, $G$ is the gravitational constant, and  the $\Sigma$ is the disk surface density. The disk should be gravitationally unstable when $Q < 1$.
If we calculate the Q parameter with the surface density profile of \citet{Cieza2018} and the {\it current} temperature profile \citep[i.e., 100 K at 39 au][]{jelee19}, it is greater than 2 at all radii and $\sim$7 at 200 au (see the red line in Figure~\ref{fig:q_parameter}). Therefore, currently, the disk appears gravitationally stable. 

The spiral structure, however, may have developed prior to the burst accretion. The disk must have a lower temperature profile before burst accretion. If the snowline was located at 4 au, then $Q < 2$ at $r < 100$ au and $Q \sim 3.5$ at 200 au (see the blue line in Figure~\ref{fig:q_parameter}).
In addition, recent works \citep{Alves2019, diaz2021} show that Class I disks with $Q > 4$ can have prominent spiral structures. 
Therefore, the spiral arms in the disk of V883 Ori could have been developed before the current outburst event. However, the spiral-like structures do not appear in the dust continuum and C$^{17}$O emission. Therefore, the spiral arms detected in the shock tracers, HCN and H$_2$CO are probably the chemical footprint that remains after the disk was stabilized.

Despite these morphological features, the velocity resolution of our data is not high enough to robustly examine the kinematics of the large arm-like structures beyond the disk. The S/N ratio of the images is not high enough to confirm the spiral structure within the disk either. Therefore, we need images with a high spatial and spectral resolution and a higher S/N ratio to inspect these features in more detail.

\subsection{Disk Rotation} \label{sec:disk_rot}

The C$^{17}$O line emission appears consistent with the smooth dust disk and is mostly confined within the disk. 
However, HCO$^+$, H$_2$CO, HCN (and its isotopologues), and HNC (and DNC) reveal a bigger disk structure than C$^{17}$O. 
\begin{figure*}[htp]
\centering
\includegraphics[width=1.0\textwidth]{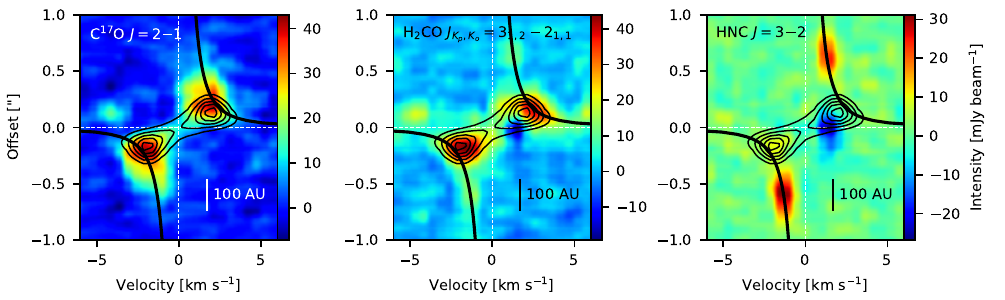}
\caption{The position-velocity (PV) diagrams for C$^{17}$O $J=$2$-$1 (left), H$_2$CO $J_{K_p,K_o}=$3$_{1,2}-$2$_{1,1}$ (middle), and HNC $J=$3$-$2 (right). The black contours represent the PV diagram of COMs lines confined within the water sublimation radius. The solid black lines indicate the Keplerian rotation profile around the central mass of 1.2 M$_{\odot}$. All PV diagrams are extracted from a line along the semi-major axis of the disk as marked in Figure~\ref{fig:mom2}. The blue-shifted blobs around -4 to -5 \kms\ between 0 to 0.2\arcsec\ are the red-shifted emissions of different adjacent lines. The negative values around 1.5 \kms\ provide evidence of infall motion; see Figure~\ref{fig:spec} for the HCO$^+$ and HCN line profiles with the infall signature, the inverse P-Cygni profile. The contours show the representative PV diagram of COMs. }
\label{fig:pv_dia}
\end{figure*}
Figure~\ref{fig:pv_dia} presents position-velocity (PV) diagrams for the C$^{17}$O (left), H$_2$CO (middle), and HNC (right) lines along the semi-major axis of the disk as marked in their moment 1 maps in Figures~\ref{fig:mom2}.  The HCN line is not used for a PV diagram along the semi-major axis because of its hyperfine structure (see the line profile in Figure~\ref{fig:mom2}). 
All three PV diagrams follow well the Keplerian rotation velocity profile, with a central mass of 1.2 M$_\odot$. 

\begin{figure*}[htp]
\centering
\includegraphics[width=0.8\textwidth]{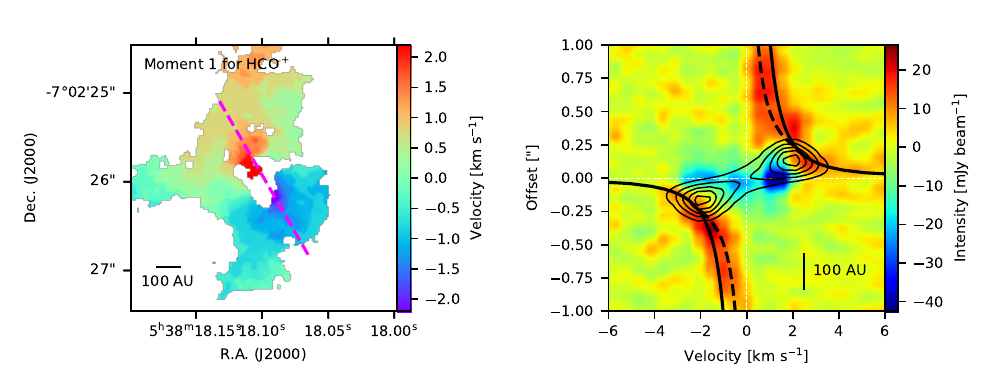}
\caption{The PV-diagrams of HCO$^+$ $J=$3$-$2. The PV diagram is constructed along the magenta dashed line in the semi-major axis (left). The black contours in the PV-diagram are the same as in Figure~\ref{fig:pv_dia}. The black solid lines show the Keplerian rotation ($1/\sqrt{r}$), while the black dashed lines indicate a rotation profile ($1/r)$ of the infalling and rotating envelope. 
The contours are the same as in Figure~\ref{fig:pv_dia}.
}
\label{fig:pv_hcop}
\end{figure*}

On the other hand, as presented in Figure~\ref{fig:pv_hcop}, the HCO$^+$ line seems to trace both the Keplerian disk (solid line) and the infalling rotating envelope (dashed line).  The large arm-like structures seen in HCO$^+$ are a hint of the connection between the envelope and the Keplerian disk (see Section \ref{subsec:large_arm_structure}). 
The black contours in all PV diagrams are made using the PCA's first Principle Component (PC1) extracted from the COMs line data \citep{Yun2023}. The PC1 provides very high S/N cube data, describing the representative spatial and velocity distribution of molecules \citep{Oko2021}. As presented in Figure~\ref{fig:coms}, all COMs emission is confined within the water sublimation radius in the disk. As a result, the most representative common feature of the COMs lines, which is the disk emission confined within the water sublimation radius, can be extracted by the PC1 of the PCA analysis of the COMs cube data.

\subsection{Infall Signature}\label{sec:infall}
\begin{figure}[htp]
\centering
\hspace{-1cm}
\includegraphics[width=0.4\textwidth]{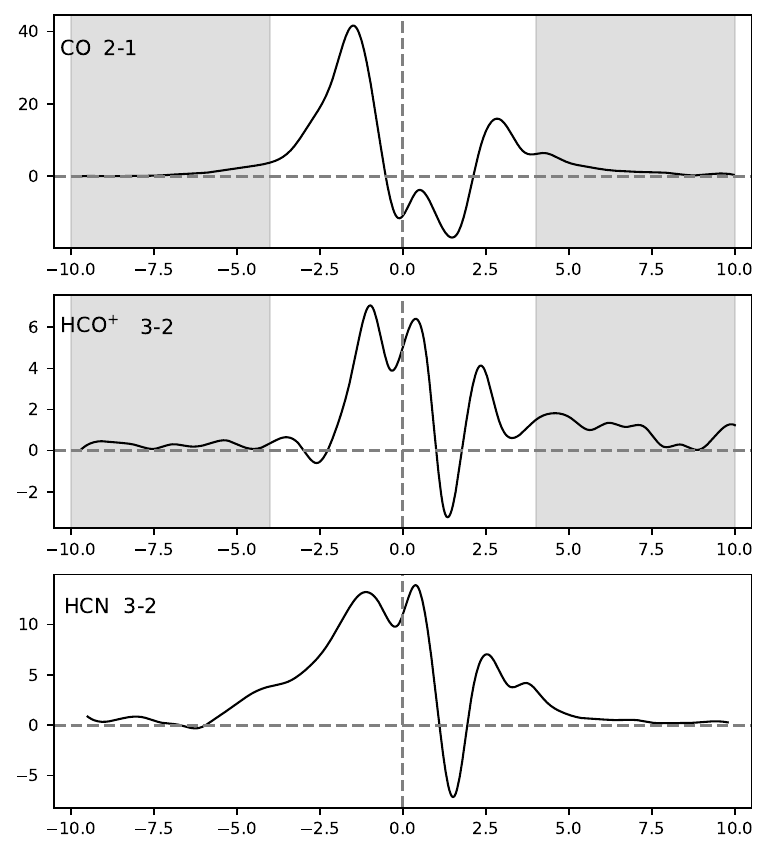}
\caption{Spectra of CO, HCO$^+$, and HCN, showing absorption below the continuum level. All spectra are extracted from a diameter of 1\mbox{$^{\prime \prime}$}. %All lines show the red-shifted absorption peaks at $\sim$1.5 \kms, indicative of infall motion.  
}
\label{fig:spec}
\end{figure}

Direct evidence of infall is provided by red-shifted absorption against the continuum \citep[inverse P-Cygni profile]{Evans15}. 
We detect the inverse P-Cygni profile in the CO, HCO$^+$, and HCN lines (Figure~\ref{fig:spec}) extracted with an aperture of 1\arcsec.
The clear absorption negative dips are seen around 1.5 \kms.
As mentioned in the previous section, in the PV diagrams of H$_2$CO and HNC, the negative features also appear at $\sim$1.5 \kms (see the middle and right panels of Figure~\ref{fig:pv_dia}). 

\begin{figure*}[htp]
\centering
\includegraphics[width=0.8\textwidth]{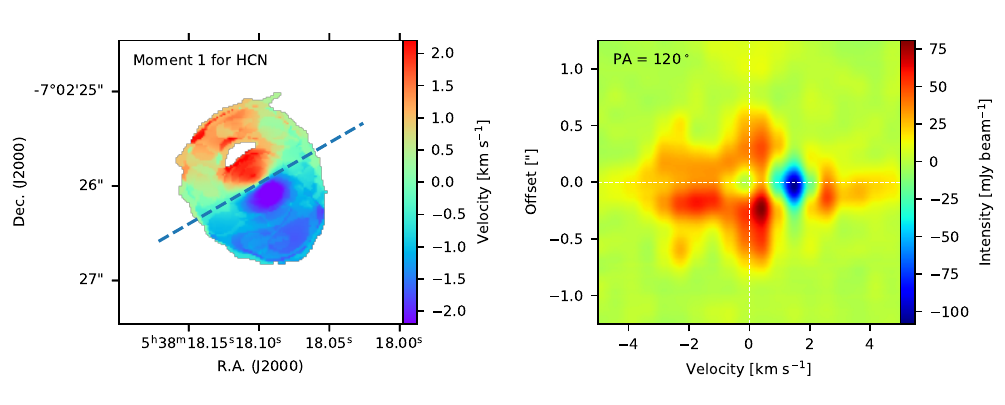}
\caption{The PV-diagram of HCN (right) along the semi-minor axis as marked with the blue dashed line at the moment 1 map (left).}% This PV-diagram has been rotated for easier comparison with that of \citet{ruiz2022}; thus, the y-axis is for velocity, unlike all other PV-diagrams in this paper.}
\label{fig:hcn_infall}
\end{figure*}
The infall features are also detected in the lower resolution ALMA observations of HCN 3--2 and HCO$^+$ 3--2 \citep{ruiz2022}. Their PV diagram of HCN along the semi-minor axis of the disk (P.A.$\sim$120\degree) shows a diamond-shape with the absorption feature, which could be reproduced by the infalling envelope within 2\arcsec\ \citep{Tobin2012}. The same feature is also shown in the PV diagram of our HCN (see Figure~\ref{fig:hcn_infall}) except for the resolution difference. 
Note that the maximum recoverable scale for the SG3 is $\sim$3.7\arcsec\, and the optically thick lines (CO, HCO$^+$, and HCN) are partially filtered out in the channel of the absorption ($\sim$1.5 \kms).

\subsection{High-Velocity Gas Components} \label{sec:hv_gas}
In Figure~\ref{fig:spec}, CO and HCO$^+$ clearly show high-velocity wings at $|v| > 4$ \kms (the gray shaded regions in the upper two panels), which is beyond the Keplerian rotation velocity expected at r $>0.1\arcsec$. 
The contours in Figure~\ref{fig:pv_dia} and \ref{fig:pv_hcop}, whose highest velocity is lower than 4 \kms, present the highest disk rotation velocity range that our observation can trace. Therefore, the high-velocity components detected in CO and HCO$^+$ are not associated with the disk rotation.
Although HCN may also trace the same high-velocity component (the bottom panel in Figure~\ref{fig:spec}), it has confusion due to its hyperfine structure.
\begin{figure*}[htp]
\centering
\hspace{0cm}
\includegraphics[width=1\textwidth]{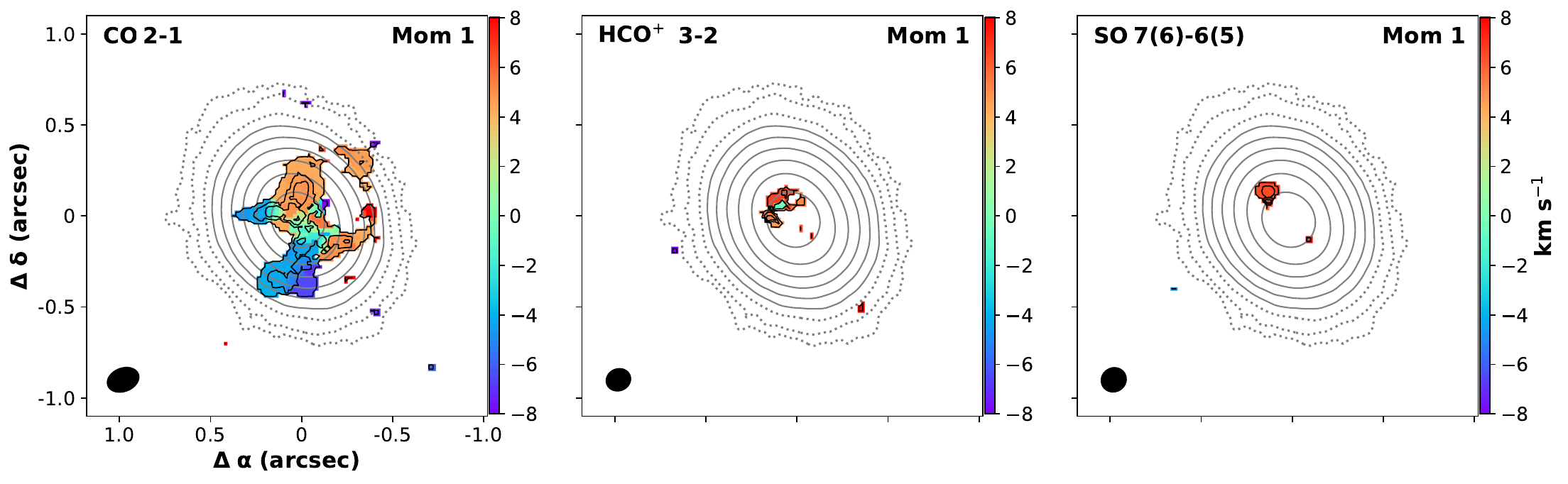}
\caption{High-velocity components. Color images are moment 1 maps of CO, HCO$^+$, and SO only for high-velocity wings ($4 < |v| < 10$ \kms). The black contour on top of color images indicates the moment 0 map of each molecule. The contour levels of the moment 0 map of CO are 6.4, 17.0, 27.6, and 38.2 mJy beam$^{-1}$ km s$^{-1}$. The contour levels of the moment 0 map of HCO$^+$ are 5.7, 11.8, 17.9, and 24.0 mJy beam$^{-1}$ km s$^{-1}$. The contour levels of the moment 0 map of SO are 4.5, 6.2, 7.9, and 9.6 mJy beam$^{-1}$ km s$^{-1}$. 
}
\label{fig:acc}
\end{figure*}

Figure~\ref{fig:acc} presents the moment 0 (contours) and 1 (images) maps of the high-velocity line components, and all of them are confined in the inner disk.
%, very close to the center.
Both red- and blue-shifted emissions for high-velocity CO components distribute along arm-like or corn-like structures, while the SO and HCO$^+$ lines show only red-shifted high-velocity components, which are concentrated in the northeast part of the disk. 

The CO velocity and morphological distribution are consistent with the larger outflow direction despite a small opening angle. We cannot, however, ignore the possibility of accreting flows. 
The SO and HCO$^+$ high-velocity emission might be caused mainly by an accretion shock since the large arm-like structures, which are traced by the SO and HCO$^+$ emission, are connected to the inner disk located inside the water sublimation radius.
%On the other hand, CN shows both red and blue high-velocity components in the north and a blue emission in the south of the inner disk. {\bf CN also shows very faint emission beyond the disk (see the third row of Figure~\ref{fig: mom5}), which might be infalling envelope material, although we need a higher S/N ratio to confirm it.}

%\begin{figure*}[htb]
%\centering
%\hspace{-1cm}
%\includegraphics[width=1.05\textwidth]{figure/new_mom/hcop_so_hnc_hcn_c17o_hdo.mom8.v4.pdf}
%\caption{Moment 8 maps of HCO$^+$, SO, HCN, HNC, C$^{17}$O, and HDO. Six different molecular emission lines tracing particular physical structures in V883 Ori: (1) HCO$^+$ traces the arm-like structure of the envelope and the largest outer disk boundary. (2) SO traces the inner disk inside the water sublimation radius and the arm-like structure in the envelope. (3) HCN traces both the outer disk ring structure and the inner disk inside the water sublimation radius. Spiral-like structures connect the outer ring and the inner disk. (4) HNC traces only the outer disk ring structure. (5) C$^{17}$O traces the gas disk consistent with the dust disk. (6) HDO traces the water sublimation radius. }
%\label{fig:overall}
%\end{figure*}

\section{A possible scenario for the current eruption in V883 Ori}\label{sec:discussion}

\begin{deluxetable*}{lll}
\tablecaption{Molecules tracing particular physical components in ALMA Band 6 \label{tb:all}}
\tablehead{
\colhead{\bf Physical component} &  \colhead{\bf Molecules}}
\startdata
Outflow cavity & CO \\
Infall signature & CO, HCO$^+$, H$_{2}$CO, HCN, HNC \\
Arm-like structure in the envelope & HCO$^+$, SO \\
Outer and inner disks connected by spiral arms & H$_{2}$CO, HCN\\
%H$^{13}$CN, HC$^{15}$N, HC$_{3}$N \\
Outer disk boundary & HNC, DNC \\
Dust disk & C$^{17}$O \\
Water sublimation radius & HDO, HDCO, HNCO, OCS, H$_2$CCO, and COMs\\
Accretion shock at inner disk edge & SO$_2$ and high-velocity wings of HCO$^+$ and SO
\enddata
\end{deluxetable*}

%Figure~\ref{fig:overall} shows six representative molecular distributions to describe the physical structures traced by different molecules and to stitch those pieces of information together to understand the burst accretion process in V883 Ori. Including those six molecules, 

Table~\ref{tb:all} summarizes the molecular tracers for various physical components detected in ALMA Band 6. 
The physical structures traced by different molecules could provide an opportunity to stitch those pieces of information together to understand the burst accretion process in V883 Ori.

The morphological (Section \ref{sec:distribution}) and kinematical (Section \ref{sec:kinematics}) phenomena related to V883 Ori, which are traced by various molecular lines, suggest that during the quiescent phase, when the disk is still cold and marginally gravitationally unstable, the infall from the envelope to the outer disk through the large arms might have induced the spiral density wave in the disk. Then, the spiral wave propagated into the central regions of the disk to trigger the accretion burst \citep{bae2014}, which, in turn, heated the disk, increasing the value of the Q parameter. Finally, the heated disk stabilized, and the spiral structures decayed. However, we still see the {\it residual} spiral structures in HCN and H$_2$CO because the timescale after the accretion burst may be much shorter than the chemical timescale within the disk, except at the dense disk midplane, which is well traced by the dust continuum and C$^{17}$O emission. As a result, the chemical footprint along the spiral structures still remains visibile on the disk.
% In the Keplerian rotation, one orbital timescale is $\sim$1000 years at $\sim$100 au around 1.2 M$_\odot$.
%This scenario is consistent with what we found in the COMs chemistry for V883 Ori; the detected and missing species of COMs in V883 Ori strongly suggest that the sublimated COMs did not have enough time to be altered by gas chemistry.

\section{Summary} \label{sec:summary}

We carried out the ALMA Spectral Survey of An eruptive Young star, V883 Ori (ASSAY) in Band 6 with a resolution of 0.15\arcsec$\sim$0.2\arcsec. First of all, HDO was detected in the disk, demonstrating that COMs are sublimated along with water. In addition, our observations reveal that different molecules trace different physical components from the infalling envelope to the Keplerian disk, where the water sublimation radius is well resolved. The combination of various molecular emission distributions illustrates a possible scenario for the burst accretion process in V883 Ori. Except for HNC and DNC, all molecules emit within the water sublimation radius ($\sim$0.3\arcsec) with an emission hole at the very central part ($<$0.1\arcsec) produced by high dust continuum opacity. 

The inner disk shows a stronger emission in the southeast half because of the disk inclination; the southeast part faces us. The combination of dust opacity and disk inclination makes the molecular emission a crescent shape. The outer boundary of the gas disk is at r$\sim$0.8\arcsec. Most molecular emission shows Keplerian rotation around a central source of 1.2 M$_\odot$, while moment 1 and 9 maps of HCO$^+$, H$_2$CO, and HCN show inflow and spiral structures.\\

The 2-D emission distributions of simple molecules are explored in this work, and the summary (from larger scales to smaller scales) is as follows:

\begin{enumerate}
%\item We detect about 4600 lines, most of which are COMs lines;
\item CO traces blue-shifted inner outflow cavity walls. The moment 1 map shows hints of inflows from envelope to disk;
\item HCO$^+$ has an emission hole larger than expected from continuum opacity alone, and that is due to its destruction by H$_2$O. As a result, HCO$^+$ traces the outer disk, producing a ring-like structure with an emission hole. 
Several arm features are connected with the ring-like structure. The velocity distribution along the arm structures is different from what is expected from the Keplerian rotation but rather consistent with the velocity of the infalling rotating envelope;
\item SO shows a compact emission inside the water sublimation radius with three weak but very prominent arm structures connected to the central emission region;
\item H$_2$CO covers the whole gas disk, but the emission inside the water sublimation radius is much stronger than the emission outside the water sublimation radius. The faint emission along the ring-like structure in the outer boundary of the disk shows spiral structures, which are similar to what is seen in HCN;
\item HCN has two distinct emission structures: one concentrates within the water sublimation radius, and the other distributes along the ring structure around the outer boundary of the dust disk. Spiral structures connect the central disk and the ring;
\item HNC and DNC present clearly a donut-shaped ring structure with missing emission within the water sublimation radius, probably due to its conversion to HCN at high temperature;
\item C$^{17}$O traces the Keplerian smooth gas disk emission without substructure, except for the emission dip at the center caused by the high dust opacity. The size of the C$^{17}$O disk is similar to the peak of HNC;
\item HDO directly proves the water sublimation radius, although H$_2$O transitions are not included within the frequency range covered by our spectral survey. The size of the HDO emission is about $\sim$0.3\arcsec, corresponding to $\sim$120 au; 
\item CH$_3$OH and other COMs also distribute only within the water sublimation radius with stronger emission in the southeast part;
\item SO$_2$ and high-velocity wings of HCO$^+$ and SO trace a shocked gas component, probably by the infalling envelope along the arm-like structure, at the northern part of the inner disk.
%\item {\bf CN} shows a strong absorption feature at the source center. The absorption line profile is symmetric respective to the source velocity and the moment 1 map shows no velocity structure, indicative of the absorption against the continuum peak by the stationary outer envelope or by a close-by interstellar cloud.
%\item {\bf HDCO} concentrate on the inner disk within the water sublimation radius, unlike their main isotopologue H$_2$CO;
\end{enumerate}

\section{Acknowledgements}

We greatly appreciate the thorough and constructive review by the anonymous referee. The review significantly improved our paper.
We appreciate Woong-Tae Kim and Jaehan Bae's helpful discussions on spiral structures. This work was supported by the National Research Foundation of Korea (NRF) grant funded by the Korean government (MSIT) (grant number 2021R1A2C1011718).
S.L.\ is supported by a Korea Astronomy and Space Science Institute grant funded by the Korean government (MSIT) (Project No. 2024-1-841-00).
G.J.H.\ is supported by general grants 12173003 and 11773002 awarded by the National Science Foundation of China.
D.J.\ is supported by NRC Canada and by an NSERC Discovery Grant. J.J.T. acknowledges support from  NSF AST-1814762.  
L.C.\ is supported by the  FONDECYT Regular program (grant No1211656) and the Millennium Nucleous YEMS (Code NCN2021\_080) awarded by the Chilean government through the ANID agency.
This research was supported by Basic Science Research Program through the National Research Foundation of Korea(NRF) funded by the Ministry of Education(grant number RS-2023-00247790).

This paper makes use of the following ALMA data: ADS/JAO.ALMA\#2019.1.00377.S.
ALMA is a partnership of ESO (representing its member states), NSF (USA), and NINS (Japan), together with NRC (Canada), NSC and ASIAA (Taiwan), and KASI (Republic of Korea), in cooperation with the Republic of Chile. The Joint ALMA Observatory is operated by ESO, AUI/NRAO, and NAOJ.

\bibliographystyle{aasjournal}
\bibliography{ms}{}

\clearpage

\clearpage
\appendix
\restartappendixnumbering
\section{Information of the observation}
\begin{figure*}[htp]
\centering
\includegraphics[width=0.5\textwidth]{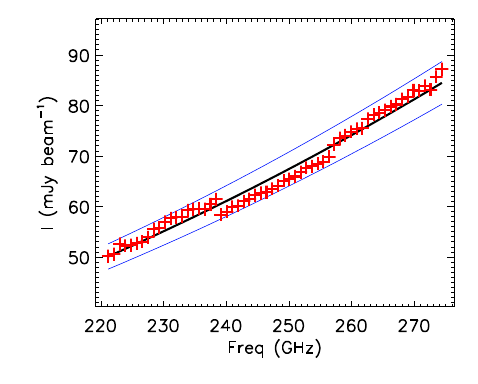}
\caption{Intensity of the averaged continuum emission toward V883 Ori for the 60 SPWs. The red crosses indicate the continuum intensity. The black solid line indicates the best fit of the frequency and intensity in the log-log scales. The blue lines indicate the absolute errors of $\pm$5 \%. The self-calibration for each Science Goal makes the continuum intensities consistent with the Science Goals, and the scatter is lower than $\pm$5 \%.  }
\label{fig:cont}
\end{figure*}

\startlongtable
\begin{deluxetable*}{ccccccc}
\tablecaption{Image parameters \label{tb:img}}
\tablehead{
\colhead{\bf SGs} & \colhead{\bf SPWs} & \colhead{\bf Frequency ranges} &\colhead{\bf b$_{\rm maj}$}&\colhead{\bf b$_{\rm min}$}&\colhead{\bf P.A.} & \colhead{\bf RMS}  \\
\colhead{} &  \colhead{} & \colhead{GHz}  &   \colhead{arcsec} &  \colhead{arcsec} &  \colhead{\degree} & \colhead{\funit} }
\startdata
\multirow{20}{*} {SG1}  &  1   &     220.70 --     221.64 & 0.15 & 0.13 & -85.2 & 2.6 \\
  &  2   &     221.61 --     222.55 & 0.16 & 0.13 & -85.2 & 2.4 \\
  &  3   &     222.52 --     223.45 & 0.17 & 0.13 & -70.8 & 2.3 \\
  &  4   &     223.42 --     224.36 & 0.15 & 0.13 & -85.9 & 2.5 \\
  &  5   &     224.33 --     225.26 & 0.17 & 0.13 & -70.9 & 2.4 \\
  &  6   &     225.23 --     226.17 & 0.17 & 0.13 & -70.4 & 2.4 \\
  &  7   &     226.14 --     227.08 & 0.16 & 0.13 & -82.6 & 2.2 \\
  &  8   &     227.05 --     227.98 & 0.16 & 0.13 & -81.9 & 2.6 \\
  &  9   &     227.95 --     228.89 & 0.18 & 0.13 & -68.4 & 2.7 \\
  & 10   &     228.86 --     229.80 & 0.18 & 0.13 & -69.0 & 2.3 \\
  & 11   &     229.77 --     230.70 & 0.18 & 0.13 & -68.7 & 2.5 \\
  & 12   &     230.67 --     231.61 & 0.18 & 0.13 & -69.8 & 2.9 \\
  & 13   &     231.58 --     232.51 & 0.18 & 0.13 & -69.1 & 2.8 \\
  & 14   &     232.48 --     233.42 & 0.18 & 0.13 & -69.1 & 2.5 \\
  & 15   &     233.39 --     234.33 & 0.18 & 0.12 & -68.9 & 2.9 \\
  & 16   &     234.30 --     235.23 & 0.18 & 0.12 & -69.0 & 2.8 \\
  & 17   &     235.20 --     236.14 & 0.17 & 0.12 & -75.9 & 3.5 \\
  & 18   &     236.11 --     237.05 & 0.19 & 0.12 & -68.7 & 2.5 \\
  & 19   &     237.02 --     237.95 & 0.19 & 0.12 & -68.8 & 3.0 \\
  & 20   &     237.92 --     238.86 & 0.19 & 0.12 & -68.8 & 3.0 \\
 \hline
\multirow{20}{*} {SG2}  &  1   &     238.70 --     239.64 & 0.22 & 0.12 & -79.7 & 2.8 \\
  &  2   &     239.61 --     240.55 & 0.22 & 0.12 & -79.9 & 2.0 \\
  &  3   &     240.52 --     241.45 & 0.22 & 0.12 & -80.1 & 2.2 \\
  &  4   &     241.42 --     242.36 & 0.22 & 0.13 & -81.1 & 2.3 \\
  &  5   &     242.33 --     243.27 & 0.22 & 0.12 & -78.6 & 2.3 \\
  &  6   &     243.23 --     244.17 & 0.23 & 0.12 & -79.0 & 2.4 \\
  &  7   &     244.14 --     245.08 & 0.22 & 0.12 & -78.8 & 2.2 \\
  &  8   &     245.05 --     245.98 & 0.22 & 0.12 & -79.8 & 2.3 \\
  &  9   &     245.95 --     246.89 & 0.23 & 0.12 & -77.6 & 2.3 \\
  & 10   &     246.86 --     247.80 & 0.23 & 0.12 & -77.9 & 2.1 \\
  & 11   &     247.77 --     248.70 & 0.23 & 0.12 & -77.9 & 2.7 \\
  & 12   &     248.67 --     249.61 & 0.23 & 0.12 & -78.9 & 2.5 \\
  & 13   &     249.58 --     250.52 & 0.24 & 0.12 & -77.3 & 3.3 \\
  & 14   &     250.48 --     251.42 & 0.23 & 0.12 & -77.4 & 2.2 \\
  & 15   &     251.39 --     252.33 & 0.23 & 0.12 & -77.1 & 2.3 \\
  & 16   &     252.30 --     253.23 & 0.23 & 0.12 & -78.0 & 2.5 \\
  & 17   &     253.20 --     254.14 & 0.23 & 0.12 & -76.2 & 2.4 \\
  & 18   &     254.11 --     255.05 & 0.23 & 0.12 & -76.2 & 2.3 \\
  & 19   &     255.02 --     255.95 & 0.23 & 0.12 & -76.2 & 2.5 \\
  & 20   &     255.92 --     256.86 & 0.23 & 0.12 & -76.9 & 2.6 \\
\hline
\multirow{20}{*} {SG3}  &  1   &     256.70 --     257.64 & 0.13 & 0.11 & -65.1 & 1.9 \\
  &  2   &     257.61 --     258.55 & 0.13 & 0.11 & -64.4 & 1.8 \\
  &  3   &     258.52 --     259.45 & 0.13 & 0.11 & -64.1 & 2.2 \\
  &  4   &     259.42 --     260.36 & 0.13 & 0.12 & -59.3 & 2.1 \\
  &  5   &     260.33 --     261.27 & 0.14 & 0.13 & -58.9 & 2.1 \\
  &  6   &     261.24 --     262.17 & 0.14 & 0.13 & -58.4 & 2.0 \\
  &  7   &     262.14 --     263.08 & 0.14 & 0.13 & -60.7 & 2.2 \\
  &  8   &     263.05 --     263.99 & 0.14 & 0.13 & -61.9 & 2.4 \\
  &  9   &     263.95 --     264.89 & 0.14 & 0.12 & -65.6 & 2.2 \\
  & 10   &     264.86 --     265.80 & 0.14 & 0.12 & -67.4 & 2.0 \\
  & 11   &     265.77 --     266.70 & 0.14 & 0.12 & -68.2 & 2.2 \\
  & 12   &     266.67 --     267.61 & 0.13 & 0.12 & -68.1 & 2.6 \\
  & 13   &     267.58 --     268.52 & 0.14 & 0.12 & -69.5 & 2.4 \\
  & 14   &     268.49 --     269.42 & 0.14 & 0.12 & -70.8 & 2.1 \\
  & 15   &     269.39 --     270.33 & 0.14 & 0.12 & -71.2 & 2.3 \\
  & 16   &     270.30 --     271.24 & 0.14 & 0.12 & -72.3 & 2.5 \\
  & 17   &     271.21 --     272.14 & 0.15 & 0.13 & -79.6 & 2.3 \\
  & 18   &     272.11 --     273.05 & 0.15 & 0.13 & -74.7 & 2.5 \\
  & 19   &     273.02 --     273.95 & 0.15 & 0.13 & -75.9 & 2.8 \\
  & 20   &     273.92 --     274.86 & 0.15 & 0.13 & -75.8 & 2.9 
\enddata
\end{deluxetable*}

\end{document}